\documentclass[12pt]{article}
\usepackage{amsmath,amssymb,graphicx}
\numberwithin{equation}{section}

\def\mydate{May 9, 2006}
\def\ignore#1{{}}

\newcounter{sxn}

\newcounter{axn}

\date{}

\newdimen\mybaselineskip
\renewcommand{\baselinestretch}{1.25}

\renewcommand{\thefootnote}{\arabic{footnote}}

\newcommand{\beeq}{\begin{equation}}
\newcommand{\eneq}{\end{equation}}
\newcommand{\beqn}{\begin{eqnarray}}
\newcommand{\eeqn}{\end{eqnarray}}

\newcommand{\alp}{\alpha}
\newcommand{\bt}{\beta}
\newcommand{\gm}{\gamma}
\newcommand{\Gm}{\Gamma}
\newcommand{\dlt}{\delta}

\newcommand{\ep}{\epsilon}
\newcommand{\tht}{\theta}

\newcommand{\lmd}{\lambda}

\newcommand{\sgm}{\sigma}

\newcommand{\vph}{\varphi}
\newcommand{\omg}{\omega}
\newcommand{\Omg}{\Omega}

\newcommand{\be}{\begin{equation}}
\newcommand{\ee}{\end{equation}}
\newcommand{\bea}{\begin{eqnarray}}
\newcommand{\eea}{\end{eqnarray}}
\newcommand{\eql}{\!\!\!&=\!\!\!&}

\newcommand{\defa}{\!\!\!&\equiv\!\!\!&}

\newcommand{\mtrx}[4]{\brkt{\begin{array}{cc}#1&#2\\#3&#4\end{array}}}

\newcommand{\vct}[2]{\brkt{\begin{array}{c}#1\\#2\end{array}}}
\newcommand{\exch}{\leftrightarrow}

\newcommand{\tl}[1]{\tilde{#1}}

\newcommand{\tr}{{\rm tr}}

\newcommand{\diag}{{\rm diag}}
\newcommand{\der}{\partial}
\newcommand{\dr}{\!\!d}
\newcommand{\hc}{{\rm h.c.}}
\newcommand{\ie}{{\it i.e.}}

\newcommand{\sgn}{{\rm sgn}}

\newcommand{\brkt}[1]{\left( #1 \right)}
\newcommand{\brc}[1]{\left\{ #1 \right\}}
\newcommand{\sbk}[1]{\left[ #1 \right]}
\newcommand{\abs}[1]{\left| #1 \right|}

\newcommand{\cD}{{\cal D}}

\newcommand{\cL}{{\cal L}}

\newcommand{\cO}{{\cal O}}
\newcommand{\cP}{{\cal P}}

\newcommand{\vb}[2]{e_{#1}^{\;\;#2}}
\newcommand{\gomg}[2]{\omega_{#1}^{\;\;#2}}
\newcommand{\zp}{z_\pi}
\newcommand{\hth}{\frac{\tht}{2}}
\newcommand{\thw}{\tht_W}
\newcommand{\hthw}{\frac{\thw}{2}}
\newcommand{\cw}{c_{\rm w}}
\newcommand{\sw}{s_{\rm w}}

\newcommand{\NP}[1]{{\it Nucl.~Phys.}~{\bf #1}}
\newcommand{\PL}[1]{{\it Phys.~Lett.}~{\bf #1}}
\newcommand{\CMP}[1]{{\it Commun.~Math.~Phys.}~{\bf #1}}

\newcommand{\PR}[1]{{\it Phys.~Rev.}~{\bf #1}}
\newcommand{\PRL}[1]{{\it Phys.~Rev.~Lett.}~{\bf #1}}
\newcommand{\PTP}[1]{{\it Prog.~Theor.~Phys.}~{\bf #1}}

\newcommand{\AP}[1]{{\it Ann.~Phys.}~{\bf #1}}

\newcommand{\RMP}[1]{{\it Rev.~Mod.~Phys }~{\bf #1}}

\def\mybig{\displaystyle \strut }

\def\dd{\partial}
\def\la{\raise.16ex\hbox{$\langle$}\lower.16ex\hbox{}  }
\def\ra{\, \raise.16ex\hbox{$\rangle$}\lower.16ex\hbox{} }
\def\go{\rightarrow}

\def\onehalf{ \hbox{${1\over 2}$} }

\def\tr{{\rm tr \,}}
\def\eff{{\rm eff}}

\def\diag{{\rm diag ~}}

\def\vphi{\varphi}
\def\ep{\epsilon}
\def\psibar{ \psi \kern-.65em\raise.6em\hbox{$-$} }
\def\psibarl{ \psi \kern-.65em\raise.6em\hbox{$-$} \lower.6em\hbox{} }

\def\myfrac#1#2{{\mybig #1\over \mybig #2}}

\begin{document}
\thispagestyle{empty}

\baselineskip=12pt

{\small \noindent \mydate    \hfill OU-HET 552/2006}


\baselineskip=35pt plus 1pt minus 1pt

\vskip 2.5cm

\begin{center}
{\Large \bf Gauge-Higgs Unification and Quark-Lepton}\\
{\Large \bf Phenomenology in the Warped Spacetime}\\


\vspace{2.0cm}
\baselineskip=20pt plus 1pt minus 1pt

{\def\thefootnote{\fnsymbol{footnote}}
\bf  Y.\ Hosotani\footnote[1]{hosotani@phys.sci.osaka-u.ac.jp},
S.\ Noda\footnote[2]{noda@het.phys.sci.osaka-u.ac.jp},
Y.\ Sakamura\footnote[3]{sakamura@het.phys.sci.osaka-u.ac.jp}
and S.\ Shimasaki\footnote[4]{shinji@het.phys.sci.osaka-u.ac.jp} }\\
\vspace{.3cm}
{\small \it Department of Physics, Osaka University,
Toyonaka, Osaka 560-0043, Japan}\\
\end{center}

\vskip 2.0cm
\baselineskip=20pt plus 1pt minus 1pt

\begin{abstract}
In the dynamical gauge-Higgs unification of electroweak interactions 
in the Randall-Sundrum warped spacetime 
the Higgs boson mass is predicted in the 
range 120 GeV -- 290 GeV, provided that the spacetime structure
is determined at the Planck scale.  Couplings of quarks and
leptons to gauge bosons and their Kaluza-Klein (KK) excited states
are determined by the masses of quarks and leptons. 
All quarks and leptons other than top quarks have very small
couplings to the KK excited states of gauge bosons.
The universality of weak interactions is slightly broken
by  magnitudes of $10^{-8}$, $10^{-6}$ and $10^{-2}$  
for $\mu$-$e$, $\tau$-$e$ and 
$t$-$e$, respectively.  Yukawa couplings become
substantially smaller than those in the standard model,
by a factor $|\cos \onehalf \theta_W|$ where $\theta_W$ is
the non-Abelian Aharonov-Bohm phase (the Wilson line phase) associated
with dynamical electroweak symmetry breaking.
\end{abstract}

\newpage


\newpage
\section{Introduction}
In the extra-dimensional gauge-Higgs unification the Higgs field 
is unified with the gauge fields so that the mass of the Higgs particle
and its self-couplings and couplings to quarks and leptons are all
determined by the underlying gauge principle and the structure of
spacetime.  As a bonus it serves as an alternative to minimal 
supersymmetric standard model to  stabilize the Higgs field 
in the electroweak interactions.  
The gauge-Higgs unification in the Randall-Sundrum (RS) 
warped spacetime has attracted much attention for its phenomenological
consequences.\cite{Pomarol2, Agashe2} 
In particular 
the Higgs mass in the dynamical gauge-Higgs unification in the RS 
warped spacetime is predicted in the energy range 120 GeV - 290 GeV 
exactly where experiments at LHC can explore.\cite{HM}  
Predictions from the gauge-Higgs unification are not limited to the Higgs sector.
The main purpose of the present paper is to show that
there appears non-universality in the weak gauge coupling 
of quarks and leptons and the Yukawa couplings of quarks and leptons 
are substantially reduced in the extra-dimensional gauge-Higgs unification
scheme.  The deviation from the universality turns out to be
small and is within the current experimental limit.  It
becomes larger for heavier leptons and quarks,  and can be tested
in future experiments. The reduction of the Yukawa couplings can be tested 
in experiments at LHC.
 
There are several key ingredients in the extra-dimensional gauge-Higgs
unification.  First, in the electroweak interactions the 
$SU(2)_L \times  U(1)_Y$ symmetry breaks down to 
$U(1)_{EM}$ by nonvanishing vacuum expectation value 
of the doublet Higgs field.  In the gauge-Higgs unification
the 4D Higgs field is identified with the extra-dimensional component 
of the gauge potentials which is necessarily in the adjoint 
representation of the gauge group.  It implies that one has to start
with a larger gauge group such as $SU(3)$, $SO(5)$, and $G_2$ to
accommodate the 4D Higgs field.  This observation was made by
Fairlie and by Forgacs and Manton.\cite{Fairlie1, Manton1}
Secondly, dynamical mechanism for electroweak symmetry breaking in the 
gauge-Higgs unification is provided by quantum dynamics of non-Abelian
Aharonov-Bohm phases (Wilson line phases), once the extra-dimensional
space is non-simply connected.\cite{YH1, YH2}  Classical vacua are
degenerate along the direction of Wilson line phases.  
The degeneracy is lifted by 
quantum effects, thereby the gauge symmetry being spontaneously broken.
This Hosotani mechanism also gives the 4D Higgs field a finite mass
by radiative corrections\cite{Lim2}.

The early attempt of dynamical gauge-Higgs unification, however, encountered severe
difficulty in incorporating chiral fermions. A major breakthrough came
in the last decade by considering an orbifold as  extra-dimensional
space.\cite{Pomarol1}  The left-right asymmetry is naturally implemented
in orbifold boundary conditions so that the matter content of the standard 
model appears in the  effective theory at low 
energies.\cite{Antoniadis1}-\cite{Panico2}  The idea has been 
applied to grand unified theory as well.\cite{Kawamura}-\cite{YH5}

One of the necessary consequences of gauge-Higgs unification is 
that the property of the Higgs particle is mostly fixed by the 
gauge principle and the structure of the extra dimension.  In particular,
the $W$ boson mass ($m_W$), the Higgs boson mass ($m_H$), and the  
Kaluza-Klein mass scale ($m_{\rm KK}$) are related to each other. 
In the dynamical gauge-Higgs unification in flat space  
one finds that $m_H \sim \sqrt{\alpha_W} \, m_W / \theta_W$ where
$\alpha_W = g_{SU(2)}^2/4\pi$ and $\theta_W$ is the Wilson line phase
(the non-Abelian Aharonov-Bohm phase) associated with the VEV of the
extra-dimensional component of gauge potentials.  
Natural matter content yields
$\theta_W$ in the range $0.2 \pi$ - $0.5 \pi$ so that $m_H$ is
found to be too small ($\sim 10$ GeV).  Further $m_{\rm KK}$  turns to be
$\sim (2\pi/\theta_W) m_W$, which also contradicts with experimental
limits.  It is not impossible to engineer a model 
 such that the resultant $\theta_W$ becomes  small enough to make
the Higgs particle sufficiently heavy, but it requires
artificial tuning of matter content.\cite{HHKY, Haba, Csaki2, Panico2}

It has been recognized that a much better and natural way of having
more realistic phenomenology in gauge-Higgs unification is to consider
gauge theory in curved spacetime.  In particular, 
in the Randall-Sundrum (RS) warped spacetime\cite{RS1} 
an enhancement factor resulting
from the spacetime curvature leads to $m_H = (120 \sim 290)$ GeV, 
just the mass value to be tested at LHC.  The Kaluza-Klein mass scale 
turns to be 1.5 TeV to 3.5 TeV as well.

The consequences of gravitational effects in the gauge-Higgs unification 
are far-reaching.  On an orbifold with topology $S^1/Z_2$ each
fermion multiplet can have its own bulk kink mass $M$.  
In the RS spacetime, in particular,
there results a natural dimensionless parameter $M/k$ where
$k$ is related to the cosmological constant $\Lambda = -k^2$
in the bulk.
These parameters are fixed by $m_W$ and quark/lepton masses, which
in turn determine wave functions of quarks and leptons in the fifth dimension.  
Couplings of quarks and leptons to gauge bosons, the Higgs
boson, and their Kaluza-Klein towers are unambiguiously 
determined.  
This procedure of calculating various couplings is previously performed 
in Ref.\cite{Agashe} although their context is not exactly the gauge-Higgs unification. 
Most of our results qualitatively reproduce their results as expected. 
However there are some consequences inherent in the gauge-Higgs unification 
in the RS spacetime. 
We will find that
the universality of the weak interactions is slightly broken,
which can be tested by future experiments.  Further
the 4D Yukawa couplings of quarks and leptons are substantially reduced.

Gauge theory in the RS spacetime has been under intensive
investigation.\cite{Rizzo}-\cite{Toms1}
The RS geometry gives a natural bridge between the Planck
scale (gravitational scale) and the weak scale by a warp factor.  
The geometry in the fifth  dimension is anti-de Sitter, 
which enables the gauge-Higgs unification in the RS spacetime 
to have  intriguing
interpretation in the AdS/CFT correspondence.\cite{Pomarol2, Agashe2}
We will see that dynamical gauge symmetry breaking 
in the RS spacetime by the Hosotani mechanism leads to intricate
structure in the electroweak interactions 
in the quark-lepton sector.  The key is the observed
quark-lepton mass spectrum, from which the fine structure in the
gauge couplings and the Higgs couplings is unambiguiously determined
for experimental verification.

In this paper we consider an $SU(3)$ gauge theory as an example. 
This model has been extensively studied in flat space for  simplicity. 
It is well known, however,  that the $SU(3)$ model does not give realistic
structure in the neutral current sector.   In particular, the Weinberg angle in 
the model turns out too big.   With this limitation in mind, we do not 
address the issue of non-universality in  neutral current interactions.
To have realistic structure in neutral current, extension of gauge group to,
say, $SO(5) \times U(1)_{B-L}$ is necessary as discussed in Ref.\cite{Agashe2}. 
It is anticipated  that most of the qualitative features obtained 
in the present paper remain valid in such modified models as well.

In section 2 gauge theory in the RS spacetime is specified with boundary
conditions. In sections 3 and 4 all fields are expanded around a nontrivial
background of the Wilson line phase, and spectra and mode functions in 
the fifth dimension are obtained.  General features of the mass spectra
are clarified in section 5.  The bulk mass parameters of quarks and leptons 
are related to their observed masses.  The mass and self-couplings of
the Higgs field are also determined there.  Nontrivial behavior of 
gauge couplings of quarks and leptons is investigated in section 6, and the
non-universality of weak interactions is established.  Reduction of Yukawa 
couplings of quarks and leptons is shown in section 7.  Section 8
is devoted to conclusion and discussions.  Useful formulae are summarized
in appendices.

\section{Gauge theory in the RS spacetime}
We consider an $SU(3)$ gauge theory on the Randall-Sundrum geometry~\cite{RS1}. 
The fifth dimension is compactified on an orbifold~$S^1/Z_2$ 
with a radius  $R$.  The bulk five-dimensional spacetime has
a negative cosmological constant $-k^2$. 
We use, throughout the paper, 
 $M,N,\cdots=0,1,2,3,4$ for the 5D curved indices, 
$A,B,\cdots=0,1,2,3,4$ for the 5D flat indices in tetrads, 
and $\mu,\nu,\cdots=0,1,2,3$  for 4D indices.\footnote{ 
As the background geometry preserves 
4D Poincar\'{e} invariance,   the curved 4D indices are not discriminated 
from the  flat 4D indices.}
The background metric is given by
\be
 ds^2 = G_{MN}dx^M dx^N
 = e^{-2\sgm(y)}\eta_{\mu\nu}dx^\mu dx^\nu+dy^2, \label{RSmetric}
\ee 
where $\eta_{\mu\nu}=\diag(-1,1,1,1)$,  $\sgm(y)=\sgm(y + 2\pi R)$, 
and $\sgm(y)\equiv k\abs{y}$ for $|y| \leq \pi R$  
with $k$ being the inverse AdS curvature radius.   

The field content consists of   $SU(3)$ gauge boson 
\be
 A_M=\sum_{a=1}^8 A_M^a \frac{\lmd^a}{2} 
 =\frac{1}{2}\brkt{\begin{array}{ccc} A^3_M+\frac{1}{\sqrt{3}}A^8_M & 
 A^1_M-iA^2_M & A^4_M-iA^5_M \\
 A^1_M+iA^2_M & -A^3_M+\frac{1}{\sqrt{3}}A^8_M & A^6_M-iA^7_M \\
 A^4_M+iA^5_M & A^6_M+iA^7_M & -\frac{2}{\sqrt{3}}A^8_M \end{array}},  
\ee
the corresponding ghost 
fields~$(\omg, \bar\omg) =\sum_a (\omg^a, \bar\omg^a) \lmd^a/2$, 
where $\lmd^a$ are the Gell-Mann matrices,  
the $SU(3)$-triplet spinors~$\psi^t = (\psi_1, \psi_2, \psi_3)$ 
and singlet spinors.  The relevant part of the action is 
\beeq
I = \int d^5 x \sqrt{-G} ~
\bigg[ \tr \Big( -\frac{1}{2}F^{MN}F_{MN}
 -\frac{1}{\xi} \, f_{\rm gf}^2
 - {\cal L}_{\omg, \bar\omg} \Big)  
+i\bar{\psi}\Gm^N \cD_N \psi -iM \ep\bar{\psi}\psi   \bigg] ~, 
 \label{cL}
\eneq
where $G\equiv\det(G_{MN}) $, $\Gm^N\equiv \vb{A}{N}\Gm^A$. 
  $\Gm^A$ is a 5D $\gamma$-matrix.
${\cal L}_{\omg, \bar\omg}$ and $M$ are the associated ghost Lagrangian and 
 a bulk mass parameter, respectively.
Since the operator~$\bar{\psi}\psi$ is $Z_2$-odd as it follows from the 
boundary condition (\ref{BC1}) below, 
we need the periodic sign function $\ep (y) = \sigma'(y)/k$
satisfying  $\ep (y) =  \pm 1$.
The field strengths and the covariant derivatives are defined by
\bea
 F_{MN} \defa \der_M A_N-\der_N A_M-ig_5\sbk{A_M,A_N} ~, \nonumber\\
 \cD_M\psi \defa \brc{\der_M-\frac{1}{4}\gomg{M}{AB}\Gm_{AB}-ig_5A_M}\psi ~, 
 \nonumber\\
 \cD_M\omg \defa \brc{\der_M-ig_5A_M}\omg ~, 
\eea
where $g_5$ is the 5D gauge coupling and 
$\Gm^{AB} = \onehalf [\Gamma^A, \Gamma^B]$. 
The spin connection 1-form $\omega^{AB} = \gomg{M}{AB} dx^M$ 
determined  from the metric~(\ref{RSmetric}) is 
\be
\omega^{\nu 4} = - \sigma' \,  e^{-\sgm} dx^\nu ~~,~~
 \mbox{other components} = 0 ~. 
\ee
The gauge-fixing function $f_{\rm gf}$ 
is specified in the next section.

The general boundary conditions in the RS spacetime are 
\bea
&&\hskip -1cm 
 A_\mu(x,-y) = P_0 A_\mu(x,y)P_0^{-1} ~~, ~~
 A_\mu(x,\pi R+y) = P_\pi A_\mu(x,\pi R-y) P_\pi^{-1} ~~, \cr
\noalign{\kern 5pt}
&&\hskip -1cm 
 A_y(x,-y) = -P_0 A_y(x,y)P_0^{-1} ~~,~~
 A_y(x,\pi R+y) = -P_\pi A_y(x,\pi R-y) P_\pi^{-1} ~~, \cr
\noalign{\kern 5pt}
&&\hskip -1cm 
 \psi(x,-y) = \eta_0 P_0 \gm_5\psi(x,y) ~~,~~
 \psi(x,\pi R+y) = \eta_\pi P_\pi\gm_5\psi(x,\pi R-y) ~~, \cr
\noalign{\kern 5pt}
&&\hskip -1cm 
 A_M(x,y+2\pi R) = UA_M(x,y)U^{-1} ~~,~~
 \psi(x,y+2\pi R) = \eta_0 \eta_\pi U\psi(x,y) ~~, 
 \label{BC1}
\eea
where $\gm_5\equiv\Gm^4$ is the 4D chiral operator and 
$\eta_0, \eta_\pi = \pm 1$. 
The unitary matrices $P_0$, $P_\pi$ and $U$ satisfy the relations
$P_0^2=P_\pi^2=1$  and $U = P_\pi P_0$. 
In the present paper  we  take $\eta_0=\eta_\pi = 1$ and
\be
 P_0 = P_\pi = \brkt{\begin{array}{ccc} -1 & & \\ & -1 & \\ & & 1 
 \end{array}}. 
 \label{BC2}
\ee
The $Z_2$-parity eigenvalues $(P_0,P_\pi)$ of  $A_M$ and $\psi$ are 
\bea
 A_\mu \eql \brkt{\begin{array}{ccc} (+,+) & (+,+) & (-,-) \\
 (+,+) & (+,+) & (-,-) \\ (-,-) & (-,-) & (+,+) \end{array}}, \;\;\;\;\;
 A_y = \brkt{\begin{array}{ccc} (-,-) & (-,-) & (+,+) \\
 (-,-) & (-,-) & (+,+) \\ (+,+) & (+,+) & (-,-) \end{array}}, \nonumber\\
 \psi_{\rm R} \eql \brkt{\begin{array}{c} (-,-) \\ (-,-) \\ (+,+) 
 \end{array}}, \;\;\;\;\;
 \psi_{\rm L} = \brkt{\begin{array}{c} (+,+) \\ (+,+) \\ (-,-) 
 \end{array}},  \label{Z2_parity}
\eea
where $\gm_5\psi_{\rm R}=\psi_{\rm R}$ 
and $\gm_5\psi_{\rm L}=-\psi_{\rm L}$. 

Note that only $(+,+)$ fields can have zero-modes when perturbation theory
is developed around the trivial configuration $A_M =0$. 
Thus the $SU(3)$ gauge symmetry is broken by the boundary condition 
to $SU(2)\times U(1)$ at the tree level. 
The zero modes of $A_y$ contain an $SU(2)$-doublet 4D scalar
$(A^4_y+iA^5_y,A^6_y+iA^7_y)$, which plays a role of the Higgs doublet 
in the standard model whose VEV breaks $SU(2)\times U(1)$ to $U(1)_{\rm EM}$.  

The zero modes of $A_y(y)$ independent of 4D coordinates $x^\mu$ 
yield non-Abelian Aharonov-Bohm phases (Wilson line phases) when integrated
along the fifth dimension.  With the residual $SU(2)\times U(1)$ 
symmetry at hand it is sufficient to consider $x^\mu$-independent zero mode
$A_y = \onehalf A_y^7 \lambda^7$, for which the Wilson line phase $\theta_W$
is given by
\beeq
  {1\over 2} \, \theta_W  = 
  g_5 \int_0^{\pi R} dy  \,  {1\over 2} \, A_y^7(y)   ~~.
\label{Wilson1}
\eneq
The factor $\onehalf$ on the left-hand side is necessary as the integral
on the right-hand side covers only a half of $S^1$.  
It has been shown in Ref.\ \cite{HM} that a gauge transformation
specified with a transformation matrix 
\beeq
\Omega(y) = \exp \bigg( i n \pi 
\myfrac{e^{2ky} - 1}{e^{2k\pi R} -1} \lambda^7 \bigg)
\quad (n: ~\hbox{an integer})
\label{largeGT1}
\eneq
preserves the boundary conditions (\ref{BC1}) and (\ref{BC2}), but
shifts the Wilson line phase by $2 n \pi$;
\beeq
\theta_W \go \theta_W' = \theta_W + 2n \pi  ~~.
\label{Wilson2}
\eneq
As a consequence $\theta_W$ is a phase variable with a period $2\pi$.

Although $\theta_W \not= 0$ gives  vanishing field strengths, it
affects physics at the quantum level.  The effective potential for
$\theta_W$ becomes non-trivial at the one loop level, whose global
minimum determines the quantum vacuum.  It is this nonvanishing 
$\theta_W$ that induces dynamical electroweak gauge symmetry breaking.
We stress that the value of $\theta_W$ is determined dynamically,
but not by hand.  Distinct boundary conditions can be equivalent
to each other at the quantum level by the dynamics of Wilson line
phases\cite{YH2}.

\section{Spectrum and mode functions of gauge bosons} \label{gauge_sct}
\subsection{General solutions in the bulk}
As the Wilson line phase $\theta_W$ acquires a nonvanishing VEV, 
we employ the background field method, 
separating $A_M$ into the classical part~$A^{\rm c}_M$ and the quantum 
part~$A^{\rm q}_M$. 
\be
 A_M = A^{\rm c}_M+A^{\rm q}_M. 
\ee
Following Oda and Weiler \cite{Oda1}, we choose
the gauge-fixing function 
\be
 f_{\rm gf} =e^{2\sgm}\eta^{\mu\nu}\cD^{\rm c}_\mu A_\nu^{\rm q}
 +\xi e^{2\sgm}\cD^{\rm c}_y\brkt{e^{-2\sgm}A^{\rm q}_y}, 
\ee
where
\be
 \cD^{\rm c}_MA^{\rm q}_N \equiv 
 \der_M A^{\rm q}_N-ig_5\sbk{A^{\rm c}_M,A^{\rm q}_N}. 
\ee
The quadratic terms for the gauge and ghost fields 
are simplified for $\xi=1$; 
\be
I = \int d^4x dy \, 
 \tr \Big[ ~ \eta^{\mu\nu}A^{\rm q}_\mu
 (\Box+\cP_4)A^{\rm q}_\nu
 + e^{-2\sgm}A^{\rm q}_y(\Box+\cP_y)A^{\rm q}_y
 +e^{-2\sgm} \bar \omg (\Box+\cP_4)\omg  ~\Big] ~, 
 \label{Lagrangian2}
\ee
where $\Box\equiv \eta^{\mu\nu}\der_\mu\der_\nu$,  
$\cP_4 \equiv \cD^{\rm c}_y e^{-2\sgm}\cD^{\rm c}_y$ and
$\cP_y \equiv \cD^{\rm c}_y\cD^{\rm c}_y e^{-2\sgm}$.
Here  we have taken 
$A_\mu^{\rm c}=0$, respecting the  4D Poincar\'{e} symmetry. 
The surface terms at the boundaries at $y=0$ and $\pi R$ 
vanish thanks to the boundary conditions for each field. 

At this stage it is most convenient to go over to the conformal
coordinate $z \equiv e^{\sgm(y)}$;
\beqn
&&\hskip -1cm
ds^2 = \myfrac{1}{z^2} \bigg\{ \eta_{\mu\nu} dx^\mu dx^\nu + 
  \myfrac{dz^2}{k^2} \bigg\} ~~, \cr
\noalign{\kern 5pt}
&&\hskip -1cm
\der_y = kz \der_z ~,~~
A_y  = kzA_z ~. 
\label{conformal1}
\eeqn
In this coordinate the boundaries are located at
$z=1$ and $\zp\equiv e^{k\pi R}$.  The action (\ref{Lagrangian2})
becomes
\beqn
&&\hskip -1cm
I = \int d^4x dz \, \myfrac{1}{kz} 
 \tr \bigg[ ~  \eta^{\mu\nu}A^{\rm q}_\mu
 (\Box+  k^2 \hat \cP_4)A^{\rm q}_\nu \cr
\noalign{\kern 5pt}
&&\hskip 2.5cm
 +  k^2 A^{\rm q}_z(\Box+ k^2 \hat \cP_z)A^{\rm q}_z
 +\frac{1}{z^2} \bar\omg (\Box+ k^2 \hat \cP_4) \omg ~\bigg] ~, \cr
\noalign{\kern 5pt}
&&\hskip -1cm
\hat \cP_4 =  z \, \cD^{\rm c}_z \,\frac{1}{z} \, \cD^{\rm c}_z ~~,~~
\hat \cP_z =  \cD^{\rm c}_z \, z \, \cD^{\rm c}_z \, \frac{1}{z} ~~.
\label{Lagrangian3}
\eeqn
The linearized equations of motion  for $A_M$ become
\bea
 \Box A^{\rm q}_\mu 
  + k^2 z \, \cD^{\rm c}_z \,\frac{1}{z} \, \cD^{\rm c}_z
  A^{\rm q}_\mu \eql 0, \nonumber\\
 \Box A^{\rm q}_z
 + k^2 \cD^{\rm c}_z \, z \, \cD^{\rm c}_z \, \frac{1}{z}
   A^{\rm q}_z
 \eql 0 ~~.  
\label{linEOM}
\eea
The classical background is taken 
to be $A_z^c = \onehalf az \lambda^7$ ($a$: constant) below.

To determine spectra and wave functions of various fields, 
we  move to a new  basis by a gauge transformation;  
\beqn
&&\hskip -1cm
\tl{A}_M \equiv \Omg A_M^{\rm q} \Omg^{-1}   ~, \cr
\noalign{\kern 10pt}
&&\hskip -1cm
\Omg(z) \equiv \exp\brc{-ig_5\int_1^z \dr z'\;A_z^{\rm c}(z')} ~. 
\label{def_Omg}
\eeqn
In the new basis the classical background of $\tl{A}_M$  vanishes 
so that $\cD_M^{\rm c}$ reduces to the simple derivative~$\der_M$, while
the boundary conditions become more involved. 
The linearized equations of motion~(\ref{linEOM}) become 
\bea
 \Box\tl{A}_\mu+k^2\brkt{\der_z^2-\frac{1}{z}\der_z}\tl{A}_\mu
 \eql 0, \nonumber\\
 \noalign{\kern 3pt}
 \Box\tl{A}_z+k^2\brkt{\der_z^2-\frac{1}{z}\der_z+\frac{1}{z^2}}\tl{A}_z 
 \eql 0. 
\eea
The  equations for eigenmodes 
with a mass eigenvalue~$m_n = k \lambda_n$ are 
\bea
\brc{\frac{d^2}{dz^2}  -\frac{1}{z}\frac{d}{dz}+\lmd_n^2}
   \tl{h}_{A,n}^a 
= \sqrt{z} \big[ - D_- (\onehalf) D_+ (\onehalf) + \lambda_n^2 \big]
  \myfrac{1}{\sqrt{z}} \, \tl{h}_{A,n}^a
\eql 0 ~~,  \cr
\noalign{\kern 5pt}
\brc{\frac{d^2}{dz^2}  -\frac{1}{z}\frac{d}{dz} +\frac{1}{z^2}
    +\lmd_n^2 }  \tl{h}_{\vph,n}^a 
= \sqrt{z} \big[ - D_+ (\onehalf) D_- (\onehalf) + \lambda_n^2 \big]
  \myfrac{1}{\sqrt{z}} \, \tl{h}_{\vph,n}^a
  \eql 0 ~~,  
\label{linEOM2}
\eea
where $D_\pm (c)$ is defind by
\beeq
D_\pm (c) \equiv \pm\,  \myfrac{d}{dz} + \myfrac{\, c\,}{z} ~~.
\label{derivative2}
\eneq
With these eigenfunctions the gauge potentials are expanded as
\be
 \tl{A}_\mu^a(x,z) = \sum_n \tl{h}^a_{A,n}(z)A_{\mu,n}(x) ~~,~~
 \tl{A}_z^a(x,z) = \sum_n \tl{h}^a_{\vph,n}(z)\vph_n(x) ~~. 
\label{expansion1}
\ee
The general solutions to Eq.(\ref{linEOM2}) are expressed 
in terms of Bessel functions as
\bea
 \tl{h}^a_{A,n}(z) \eql z\brc{\alp^a_{A,n}J_1(\lmd_n z)
 +\bt^a_{A,n}Y_1(\lmd_n z)}, \nonumber\\
 \tl{h}^a_{\vph,n}(z) \eql z\brc{\alp^a_{\vph,n}J_0(\lmd_n z)
 +\bt^a_{\vph,n}Y_0(\lmd_n z)},  
 \label{gen_sol1}
\eea
where $\alp^a_{A,n}$, $\bt^a_{A,n}$, $\alp^a_{\vph,n}$ and 
$\bt^a_{\vph,n}$ are constants to be determined.

\subsection{Mass eigenvalues and mode functions}
To determine the eigenvalues $\lambda_n$'s and the corresponding
mode functions (\ref{gen_sol1}),  we need to take into account 
the boundary conditions (\ref{BC1}) and (\ref{BC2}). 
It follows from the action (\ref{Lagrangian2}) or 
(\ref{Lagrangian3}) that 
$\tr A_\mu \dd_z A^\mu$ and $\tr A_z \dd_z (A_z/z)$ must
vanish at $z=1$ and $z_\pi$. 
For $Z_2$-even components in (\ref{Z2_parity}), therefore, one has
\beeq
\dd_z A_\mu^a  =0 ~~,~~
\dd_z \Big( \frac{1}{z} A_z^a \Big)  = 0 
\qquad {\rm at}~ z=1, z_\pi ~,
\label{bd_even}
\eneq
while for $Z_2$-odd components, 
\beeq
A_\mu^a = A_z^a = 0 \qquad {\rm at}~ z=1, z_\pi ~.
\label{bd_odd}
\eneq
One has to translate the  conditions~(\ref{bd_even}) 
and (\ref{bd_odd}) into those in the new basis $\tl{A}_M$, or
for (\ref{expansion1}) and (\ref{gen_sol1}).

As is inferred from (\ref{linEOM}), $Z_2$-even components of $A_z$
have zero modes ($\lambda_0=0$) with $A_z \propto z$.
Making use of the residual $SU(2) \times U(1)$ symmetry, we
can restrict ourselves to
\beeq
A_z^{\rm c} = \frac{1}{2} az \lambda^7 
\label{background2}
\eneq
where $a$ is an arbitrary constant. The constant $a$ is related to
$\theta_W$ in (\ref{Wilson1}) by
\beeq
\theta_W = \frac{1}{2} \, g_5 a (z_\pi^2 -1) ~.
\label{Wilson3}
\eneq
The potential has  a classical flat direction along $\theta_W$.
The value for $\theta_W$ is determined at the  quantum level.
The  gauge transformation matrix~$\Omg$ 
defined in Eq.(\ref{def_Omg}) becomes
\be
 \Omg(z) = \exp\brc{-i \onehalf \tht(z)\lmd^7} 
 = \brkt{\begin{array}{ccc} 1 & & \\ & \cos \onehalf \tht & -\sin \onehalf\tht \\
 & \sin \onehalf \tht & \cos \onehalf \tht \end{array}}, 
 \label{expr_Omg}
\ee
where 
\be
 \tht(z) \equiv g_5\int_1^z\dr z'\;A^7_z(z')
  = \frac{g_5a}{2} (z^2-1) 
  = \theta_W ~ \myfrac{z^2-1}{z_\pi^2 -1} ~.
\label{transformation2}
\ee
Thus the relation between $A_M$ and $\tl{A}_M$ in (\ref{def_Omg}) 
can be written as 
\bea
 \vct{\tl{A}^1_M}{\tl{A}^4_M} \eql 
 \mtrx{\cos\onehalf\tht}{-\sin\onehalf\tht}{\sin\onehalf\tht}{\cos\onehalf\tht}
 \vct{A^1_M}{A^4_M}, \nonumber\\
\noalign{\kern 5pt}
 \vct{\tl{A}^2_M}{\tl{A}^5_M} \eql 
 \mtrx{\cos\onehalf\tht}{-\sin\onehalf\tht}{\sin\onehalf\tht}{\cos\onehalf\tht}
 \vct{A^2_M}{A^5_M}, \nonumber\\
\noalign{\kern 5pt}
 \vct{\tl{A}^{'3}_M}{\tl{A}^6_M} \eql 
 \mtrx{\cos\tht}{-\sin\tht}{\sin\tht}{\cos\tht}
 \vct{A^{'3}_M}{A^6_M}, \nonumber\\
\noalign{\kern 5pt}
 \tl{A}^7_M \eql A^7_M ~~,~~
 \tl{A}^{'8}_M = A^{'8}_M, 
 \label{rel_tlA-A}
\eea
where 
\be
 \vct{A^{'3}_M}{A^{'8}_M} \equiv \mtrx{-\frac{1}{2}}{\frac{\sqrt{3}}{2}}
 {-\frac{\sqrt{3}}{2}}{-\frac{1}{2}} \vct{A^3_M}{A^8_M}. 
\ee

For $(A^1_\mu,A^4_\mu)$, for example, 
the boundary conditions~(\ref{bd_even}) and (\ref{bd_odd}) become 
\bea
\frac{d}{dz} \brkt{\cos\hth\cdot\tl{h}^1_{A,n}
 +\sin\hth\cdot\tl{h}^4_{A,n}} \bigg|_{z=1,z_\pi} \eql 0 ~, \cr
\noalign{\kern 5pt}
-\sin\hth\cdot\tl{h}^1_{A,n}
 +\cos\hth\cdot\tl{h}^4_{A,n} \bigg|_{z=1,z_\pi} \eql 0 ~.
 \label{BC4}
\eea
The conditions are summarized as
\be
 \brkt{\begin{array}{cccc} \lmd_n J_0(\lmd_n) & \lmd_n Y_0(\lmd_n) 
 & 0 & 0 \\ \cw\lmd_n J_0(\lmd_n\zp) & \cw\lmd_n Y_0(\lmd_n\zp) 
 & \sw\lmd_n J_0(\lmd_n\zp) & \sw\lmd_n Y_0(\lmd_n\zp) \\
 0 & 0 & J_1(\lmd_n) & Y_1(\lmd_n) \\
 -\sw J_1(\lmd_n\zp) & -\sw Y_1(\lmd_n\zp) & \cw J_1(\lmd_n\zp) 
 & \cw Y_1(\lmd_n\zp) \end{array}}
 \brkt{\begin{array}{c} \alp^1_{A,n} \\ \bt^1_{A,n} \\
 \alp^4_{A,n} \\ \bt^4_{A,n} \end{array}} = 0 ~, 
 \label{cond_alp_bt}
\ee
where $\cw \equiv \cos \onehalf \theta_W$ and 
$\sw \equiv \sin \onehalf \theta_W$. 
For a nontrivial solution   to exist, 
the determinant of the above $4\times 4$ matrix must vanish,
which leads to
\be
 \lmd_n^2\zp F_{0,0}(\lmd_n,\zp)F_{1,1}(\lmd_n,\zp) 
 = \frac{4}{\pi^2}\sin^2\hthw  ~~.
 \label{detM}
\ee
Here  $F_{\alp,\bt}(\lmd,z)$ is defined 
in (\ref{def_F}) in Appendix~\ref{fml}. 
Eq.\ (\ref{detM})  determines the  eigenvalues~$\lmd_n$. 
Once $\lambda_n$ is determined, the corresponding $\alpha_n$'s 
and $\beta_n$'s are fixed by (\ref{cond_alp_bt}) with 
the normalization conditions\footnote{
Due to the twisting by $\Omg(z)$, each mode has nonzero components  
in both $\tl{A}^1_\mu$ and $\tl{A}^4_\mu$. } 
\beeq
\int_1^{\zp} dz\, \frac{1}{kz}
\brc{\tl{h}^1_{A,n}(z)\tl{h}^1_{A,l}(z)
+\tl{h}^4_{A,n}(z)\tl{h}^4_{A,l}(z)} = \dlt_{nl} ~.
\label{normalization1}
\eneq
The result is 
\bea
\tl{h}^1_{A,n} (z) \eql 
C^{\rm d}_{1,n}(\thw)\cdot zF_{1,0}(\lmd_n,z) ~, \cr
\noalign{\kern 10pt} 
\tl{h}^4_{A,n}(z) \eql 
 C^{\rm s}_{1,n}(\thw)\cdot zF_{1,1}(\lmd_n,z) ~,
\label{wavefunction1}
\eea
where the coefficients~$C^{\rm d,s}_{\alp,n}(\thw)$ are defined 
in (\ref{def_F_Cs}).   Similarly, one finds, for $(A^1_z, A^4_z)$, that
\beqn
&&\hskip -1cm
\int_1^{\zp} dz\, \frac{k}{~z~}
\brc{\tl{h}^1_{\vphi,n}(z)\tl{h}^1_{\vphi,l}(z)
+\tl{h}^4_{\vphi,n}(z)\tl{h}^4_{\vphi,l}(z)} = \dlt_{nl} ~,  \cr
\noalign{\kern 5pt}
&&\hskip -1cm
\tl{h}^1_{\vphi,n}(z) =
\myfrac{1}{k}~ C^{\rm d}_{1,n}(\thw)\cdot   zF_{0,0}(\lmd_n,z) ~,  \cr
\noalign{\kern 0pt}
&&\hskip -1cm
\tl{h}^4_{\vphi,n}(z) =
\myfrac{1}{k}~ C^{\rm s}_{1,n}(\thw) \cdot   zF_{0,1}(\lmd_n,z) ~.
\label{wavefunction1b}
\eeqn

From Eqs.(\ref{Z2_parity}) and (\ref{rel_tlA-A}), we can see that 
the same formulae are obtained for $(A^2_\mu,A^5_\mu)$;
$\tl{h}^2_{A,n}(z)=\tl{h}^1_{A,n}(z)$,
$\tl{h}^5_{A,n}(z)=\tl{h}^4_{A,n}(z)$, etc.
The lightest mode in $(A^1_\mu+iA^2_\mu,A^4_\mu+iA^5_\mu)$ is 
the $W$ boson for the electroweak interactions. 

We remark that $A_\mu^a$ and $A_z^a$ have a degenerate mass spectrum 
except for the zero-mode. 
Differentiating  the first equation in Eq.(\ref{linEOM2}) 
with respect to $z$,  one finds that
\be
\frac{d}{dz} \brc{ \frac{d^2}{dz^2} 
-\frac{1}{z} \frac{d}{dz}+\lmd_n^2}\tl{h}^a_{A,n} 
 =\brc{ \frac{d^2}{dz^2} 
 -\frac{1}{z} \frac{d}{dz} +\frac{1}{z^2} + \lmd_n^2 }
 \frac{d \tl{h}^a_{A,n}}{dz} =0 ~. 
\ee
Comparing it with the second equation in Eq.(\ref{linEOM2}), 
we observe that $d \tl{h}^a_{A,n}/dz$ satisfies the same 
mode equation as  $\tl{h}^a_{\vph,n}(z)$ does.
Since $d \tl{h}^a_{A,n}/dz$ and $\tl{h}^a_{\vph,n}(z)$ satisfy
 the same boundary condition,  
$d \tl{h}^a_{A,n}/dz \propto \tl{h}^a_{\vph,n}$  
and the corresponding modes~$A_{\mu,n}(x)$ and $\vph_n(x)$ have 
the same eigenvalue~$\lmd_n$.\footnote{ 
This correspondence holds only for nonzero-modes 
since the mode function for the zero-mode of $\tl{A}^a_\mu$ is a constant.}

\section{Spectrum and mode functions of fermions}
 \label{ferm_sct}
Next we consider the fermion sector. 
From the action (\ref{cL}), the linearized equation of motion is 
\be
 i\Gm^N\brkt{\der_N-\frac{1}{4}\gomg{N}{AB}\Gm_{AB}-ig_5A^{\rm c}_N}\psi
 -iM \ep\psi = 0 ~~. 
 \label{fermionEq1}
\ee
Let us restrict ourselves to the fundamental region
 $0\leq y \leq\pi$ or $1 \le z \le z_\pi$ where $\ep=1$. 
As in the previous section, the Kaluza-Klein decomposition becomes easier
in the new basis (\ref{def_Omg}).  We introduce  $\tilde \psi$ by
\be
 \tl{\psi} \equiv z^{-2}\Omg(z)\psi,  \label{rel_tlps-ps}
\ee
with $\Omg(z)$  defined in (\ref{def_Omg}). 
Then, Eq.\ (\ref{fermionEq1}) becomes 
\bea
 \gm^\mu\der_\mu\tl{\psi}_{\rm R}-\brkt{k\der_z+\frac{M}{z}}\tl{\psi}_{\rm L} 
 \eql 0,  \nonumber\\
 \gm^\mu\der_\mu\tl{\psi}_{\rm L}-\brkt{-k\der_z+\frac{M}{z}}\tl{\psi}_{\rm R}
 \eql 0, \label{ferm_linEOM}
\eea
where $\gm^\mu$ is the 4D $\gm$-matrices 
defined by $\gm^\mu\equiv\Gm^{A=\mu}$.\footnote{
Note that $\Gm^{M=\mu}=e^\sgm\gm^\mu$. 
Throughout the paper,  4D indices are raised and lowered by $\eta^{\mu\nu}$ 
and $\eta_{\mu\nu}$, respectively.}
Here we have decomposed $\tl{\psi}$ into the eigenstates of $\gm_5$, \ie, 
$\tl{\psi}=\tl{\psi}_{\rm R}+\tl{\psi}_{\rm L}$ where 
$\gm_5\tl{\psi}_{\rm R}=\tl{\psi}_{\rm R}$ 
and $\gm_5\tl{\psi}_{\rm L}=-\tl{\psi}_{\rm L}$. 
From these equations, the mode equations for the fermion are given by
\be
 D_\pm \Big( \frac{M}{k} \Big) \tl{f}^\mp_{i,n} (z)
 = -\lmd_n \tl{f}^\pm_{i,n} (z) ~, 
\label{ferm_md_eq}
\ee
where $i$ is an $SU(3)$-triplet index and $D_\pm(c)$ is defined 
in (\ref{derivative2}).  $\tl{\psi}_{{\rm R}i}$ and $\tl{\psi}_{{\rm L}i}$
are expaned as 
\be
 \tl{\psi}_{{\rm R}i}(x,z) = \sum_n\tl{f}^+_{i,n}(z)\psi_n^+ (x) ~~,~~  
 \tl{\psi}_{{\rm L}i}(x,z) = \sum_n\tl{f}^-_{i,n}(z)\psi_n^- (x) ~~. 
 \label{expansion2}
\ee
The general solutions to Eq.(\ref{ferm_linEOM}) are 
\bea
 \tl{f}^+_{i,n}(z) \eql z^{\frac{1}{2}}
  \big\{ a^+_{i,n} J_{\alp-1}(\lmd_n z) 
 +b^+_{i,n} Y_{\alp-1}(\lmd_n z) \big\} ~, \nonumber\\
\noalign{\kern 5pt}
 \tl{f}^-_{i,n}(z) \eql z^{\frac{1}{2}}
 \big\{ a^-_{i,n} J_\alp(\lmd_n z) 
 +b^-_{i,n} Y_\alp(\lmd_n z) \big\} ~, 
\label{fermi_mode1}
\eea
where $\alp\equiv (M/k) +\frac{1}{2}$. 
The eigenvalue $\lambda_n$ and the coefficients $a^\pm_{i,n}$, $b^\pm_{i,n}$
are determined by the boundary conditions. 

To figure out the boundary conditions for $\tilde \psi$,  we look at the 
action and equations in the original basis. Taking into account the fact that 
$A_y$ is continuous at the boundaries, one finds that
$\hat \psi = z^{-2} \psi = \Omega^{-1} \tilde \psi$ must obey
\beqn
&&\hskip -1cm
D_+ \hat\psi_{{\rm L}1} = D_+ \hat\psi_{{\rm L}2} 
= D_- \hat\psi_{{\rm R}3} = 0 ~~,  \cr
\noalign{\kern 5pt}
&&\hskip -1cm
\hat\psi_{{\rm R}1} = \hat\psi_{{\rm R}2} = \hat\psi_{{\rm L}3} = 0 ~~, 
\label{BCf1}
\eeqn
at $z = 1$ and $z_\pi$, where $D_\pm = D_\pm (M/k)$. 
$\tilde \psi$ is related to $\hat \psi$ by 
\bea
\hat \psi_1 \eql  \tl{\psi}_1 ~~, \cr
\noalign{\kern 10pt}
\hat \psi_2 \eql  \brc{\cos\frac{\tht(z)}{2}\cdot\tl{\psi}_2
 +\sin\frac{\tht(z)}{2}\cdot\tl{\psi}_3} ~~, \cr
\noalign{\kern 5pt}
\hat \psi_3 \eql  \brc{-\sin\frac{\tht(z)}{2}\cdot\tl{\psi}_2
 +\cos\frac{\tht(z)}{2}\cdot\tl{\psi}_3} ~~. 
\label{relation2}
\eea
$\tilde \psi_1$ is expanded in modes by itself, 
while $\tilde \psi_2$ and $\tilde \psi_3$ are expanded in 
a single KK tower,  each mode of which has nonvanishing support 
on both $\tilde \psi_2$ and $\tilde \psi_3$ 
for $\thw\neq 0$ mod $2\pi$. Taking this fact into account, 
 we label the KK modes in 
$\tilde \psi_1$ and $(\tilde \psi_2, \tilde \psi_3)$ separately. 

Mode functions are obtained in the same way as in the case 
of the gauge fields. Normalization conditions are given by
\bea
&&\hskip -1cm
\int_1^{\zp} \frac{dz}{k} \tl{f}^\pm_{1,n}(z)\tl{f}^\pm_{1,l}(z) 
  = \dlt_{nl} ~~, \cr
\noalign{\kern 5pt}  
&&\hskip -1cm
\int_1^{\zp} \frac{dz}{k} \brc{ \tl{f}^\pm_{2,n}(z)\tl{f}^\pm_{2,l}(z)
 +\tl{f}^\pm_{3,n}(z)\tl{f}^\pm_{3,l}(z) } = \dlt_{nl}~~.
\label{normalization2} 
\eea
For the right-handed components 
\bea
\tl{f}^+_{1,l}(z) \eql 
\begin{cases}
~ 0 &\hbox{for }l=0~, \cr
 \myfrac{\sqrt{k}\pi \lmd^0_l}{\sqrt{2}} 
 \brc{\myfrac{Y_{\alp-1}^2(\lmd^0_l)}{Y_{\alp-1}^2(\lmd^0_l\zp)}-1}^{-\frac{1}{2}}
 \cdot z^{\frac{1}{2}}F_{\alp-1,\alp-1}(\lmd^0_l,z) &\hbox{for } l\neq 0 ~,
\end{cases}  
 \cr
\noalign{\kern 10pt}
\tl{f}^+_{2,n}(z) \eql 
 \, C^{\rm d}_{\alp,n}(\thw)\cdot
 z^{\frac{1}{2}}F_{\alp-1,\alp-1}(\lmd_n,z) ~~, \cr
\noalign{\kern 10pt}
\tl{f}^+_{3,n}(z) \eql 
\, C^{\rm s}_{\alp,n}(\thw)\cdot
 z^{\frac{1}{2}}F_{\alp-1,\alp}(\lmd_n,z) ~~,  
\label{RH_md_fn}
\eea
whereas for the left-handed components 
\bea
\tl{f}^-_{1,l}(z) \eql
\begin{cases}
\brkt{ \myfrac{2k(1-\alp)}{\zp^{2(1-\alp)}-1}}^{\frac{1}{2}}
  \cdot z^{\frac{1}{2}-\alp} &\hbox{for }l=0~, \cr
 \myfrac{\sqrt{k}\pi\lmd^0_l}{\sqrt{2}}
 \brc{\myfrac{Y_{\alp-1}^2(\lmd^0_l)}{Y_{\alp-1}^2(\lmd^0_l\zp)}-1}^{-\frac{1}{2}}
 \cdot z^{\frac{1}{2}}F_{\alp,\alp-1}(\lmd^0_l,z) &\hbox{for } l\neq 0 ~,
\end{cases}  
 \cr
\noalign{\kern 10pt}
\tl{f}^-_{2,n}(z) \eql 
 \, C^{\rm d}_{\alp,n}(\thw)\cdot 
 z^{\frac{1}{2}}F_{\alp,\alp-1}(\lmd_n,z) ~~, \cr
\noalign{\kern 10pt}
\tl{f}^-_{3,n}(z) \eql 
 \, C^{\rm s}_{\alp,n}(\thw)\cdot
 z^{\frac{1}{2}}F_{\alp,\alp}(\lmd_n,z) ~~.
\label{LH_md_fn}
\eea
The functions~$F_{\alp,\bt}(\lmd_n,z)$ and $C^{\rm d,s}_{\alp,n}(\thw)$ 
are defined in (\ref{def_F}) and (\ref{def_F_Cs}).  
The  eigenvalues~$\lmd_l^0$ and $\lmd_n$ are the solutions of 
\bea
 F_{\alp-1,\alp-1}(\lmd_l^0,\zp) \eql 0 ~, \label{detM_ferm1} \\
\noalign{\kern 10pt}
 \lmd_n^2\zp F_{\alp-1,\alp-1}(\lmd_n,\zp)F_{\alp,\alp}(\lmd_n,\zp) 
 \eql \frac{4}{\pi^2}\sin^2\hthw ~,  \label{detM_ferm2}
\eea
respectively. 
Only $\psi_{{\rm L}1}$ has a zero-mode if $\thw \not= 0$ mod $2\pi$.  

Note that the left- and the right-handed modes have 
degenerate mass eigenvalues for each KK level except for the zero-mode, 
as inferred  from Eq.(\ref{ferm_md_eq}). 
\ignore{There is one-to-one correspondence between the left- 
and the right-handed modes with the same KK level $n ~(\neq 0)$. } 
It is easy to show that $(a^+_{i,n},b^+_{i,n})=(a^-_{i,n},b^-_{i,n})$ 
in (\ref{fermi_mode1}) for $\lambda_n \not= 0$.   With the aid of 
(\ref{flip_F}) and (\ref{flip_C}), one can see that 
the mode functions satisfy, 
under the flip of the sign of the bulk mass~$M\exch -M$,  that
\bea
 \tl{f}^+_{2,n}(z) & \exch & p_{\alp,n}(\thw)\tl{f}^-_{3,n}(z) ~, \nonumber\\
 \tl{f}^+_{3,n}(z) & \exch & -p_{\alp,n}(\thw)\tl{f}^-_{2,n}(z) ~,  
\label{flip_md_fn}
\eea
where the sign factor~$p_{\alp,n}(\thw) = \pm 1$ is defined by (\ref{def_p}). 

In passing we would like to comment that the spectra and wave functions
of various fields in the RS spacetime reveal the structure of 
supersymmetric (SUSY) quantum mechanics.  
The pair $(\tilde h^a_{A,n}, \tilde h^a_{\vphi,n})$ for gauge fields
and the sets of pairs $(\tilde f^+_{i,n}, \tilde f^-_{i,n})$ for
fermions form bases for the SUSY structure.  Eqs.\ (\ref{linEOM2}) and
(\ref{ferm_md_eq}) with the designated boundary conditions 
guarantee quantum mechanics SUSY.  This feature for gauge fields
has been stressed in Ref.\ \cite{Lim3} in general 5D warped
spacetime.

\section{Mass spectrum}
\subsection{General properties of mass spectrum} 
\label{pms}

As we have seen in the previous two sections, 
there are two types of fields with respect to their KK decomposition. 
The first is of the singlet-type which is unrotated by $\Omg(z)$ in
(\ref{expr_Omg}).  Its KK spectrum is not affected by the nonvanishing
Wilson line phase~$\thw$. 
The other is of the doublet-type which is rotated by $\Omg(z)$.  
The KK spectrum of the latter type of fields  depend on $\thw$. 
In this section we investigate the $\thw$-dependence of their KK spectrum. 

The mass spectrum~$\brc{m_n = k\lmd_n}$ of fields of the doublet-type
is determined by
\be
 \lmd_n^2\zp F_{\alp-1,\alp-1}(\lmd_n,\zp)F_{\alp,\alp}(\lmd_n,\zp) 
 = \frac{4}{\pi^2}\sin^2\hthw.  \label{mass_det}
\ee
$\alp=1$ for gauge fields, while $\alp=(M/k)+\frac{1}{2}$ 
for fermion fields.\footnote{
For $(A^{'3}_M,A^6_M)$, $\thw$ in Eq.\ (\ref{mass_det}) is replaced 
by $2\thw$. }
(See Eqs.\ (\ref{detM}) and (\ref{detM_ferm2}).)
Using Eq.(\ref{rel_for_F}), the equation can also be written as 
\be
 \lmd_n^2\zp F_{\alp-1,\alp}(\lmd_n,\zp)F_{\alp,\alp-1}(\lmd_n,\zp)
 = -\frac{4}{\pi^2}\cos^2\hthw. \label{mass_det2}
\ee 

As confirmed by numerical evaluation of (\ref{mass_det}), 
the smallest mass eigenvalue satisfies $\lmd_0\zp\ll 1$ 
when the warp factor~$\zp=e^{k\pi R}$ is large enough. 
Making use of (\ref{limit_F}), 
one finds, for  $m_0=k\lmd_0$, 
\be
 m_0 = 
 k\brkt{\frac{\alp(\alp-1)}
 {\zp\sinh(\alp k\pi R)\sinh((\alp-1)k\pi R)}}^{\frac{1}{2}}
  \Big| \sin \frac{\theta_W}{2} \Big|
 \brc{1+\cO\brkt{\frac{m_0^2\zp^2}{k^2}}}.  \label{ap_m0}
\ee
The correction terms are of order $(\pi m_0/m_{\rm KK})^2$ 
for $\zp\gg 1$, where 
\be
 m_{\rm KK} \equiv \frac{k\pi}{\zp-1} \label{def_mKK}
\ee
is the  KK mass scale. 
In particular, the mass of the $W$ boson is given by 
\be
 m_W = \frac{m_{\rm KK}}{\pi}
 \brkt{\frac{2}{\pi kR}}^{\frac{1}{2}}  
 \Big| \sin \frac{\theta_W}{2} \Big|
 \brc{1+\cO\brkt{\frac{\pi^2 m_W^2}{m_{\rm KK}^2}}},  \label{ap_mW}
\ee
for $\zp\gg 1$. 
Note that the formula (\ref{ap_mW}) is consistent with the result 
in Ref.~\cite{HM} for $\thw\ll 1$. 
In the flat spacetime ($k=0$), the correction terms are no longer 
negligible in (\ref{ap_m0}). 
As will be seen, this modification amounts to the replacement 
$\sin\onehalf \thw \to \onehalf \thw$ in (\ref{ap_m0}). 

One can draw an important consequence from (\ref{ap_mW}). The RS
spacetime is specified with two parameters, $k$ and $kR$.  If one
supposes that the structure of spacetime is determined at the Planck
scale so that $k \sim M_{\rm Pl}$, then (\ref{ap_mW}) implies that
$kR = 12 \pm 0.1$.
It has been known \cite{HHKY, HNT2, HHHK, Lim1, HNT1}
that with natural matter content
the effective potential $V_\eff(\theta_W)$  has a global minimum
either at $\theta_W=0$, at $\theta_W = (0.2 \sim 0.8) \pi$ 
or at $\theta_W=\pi$,
the second of which corresponds to the electroweak symmetry breaking.
Therefore the value of $kR$ is determined around 12 irrespective
of the details of the model considered.  We take $kR=12$ in 
the numerical evaluation in the rest of the paper.

Once $kR$ and $m_W$ are given, $m_{\rm KK}$ is determined 
as a function of $\theta_W$.  
One sees that $m_{\rm KK} = 1.5 \,$TeV $\sim$ $3.5 \,$TeV
for $\theta_W = \onehalf\pi$ $\sim$ $\frac{1}{5}\pi$.  
$(\onehalf \pi kR)^{1/2}$ gives an important enhancement factor
in the RS spacetime.

\begin{figure}[t,h,b]
\centering  \leavevmode
\includegraphics[width=5.2cm]{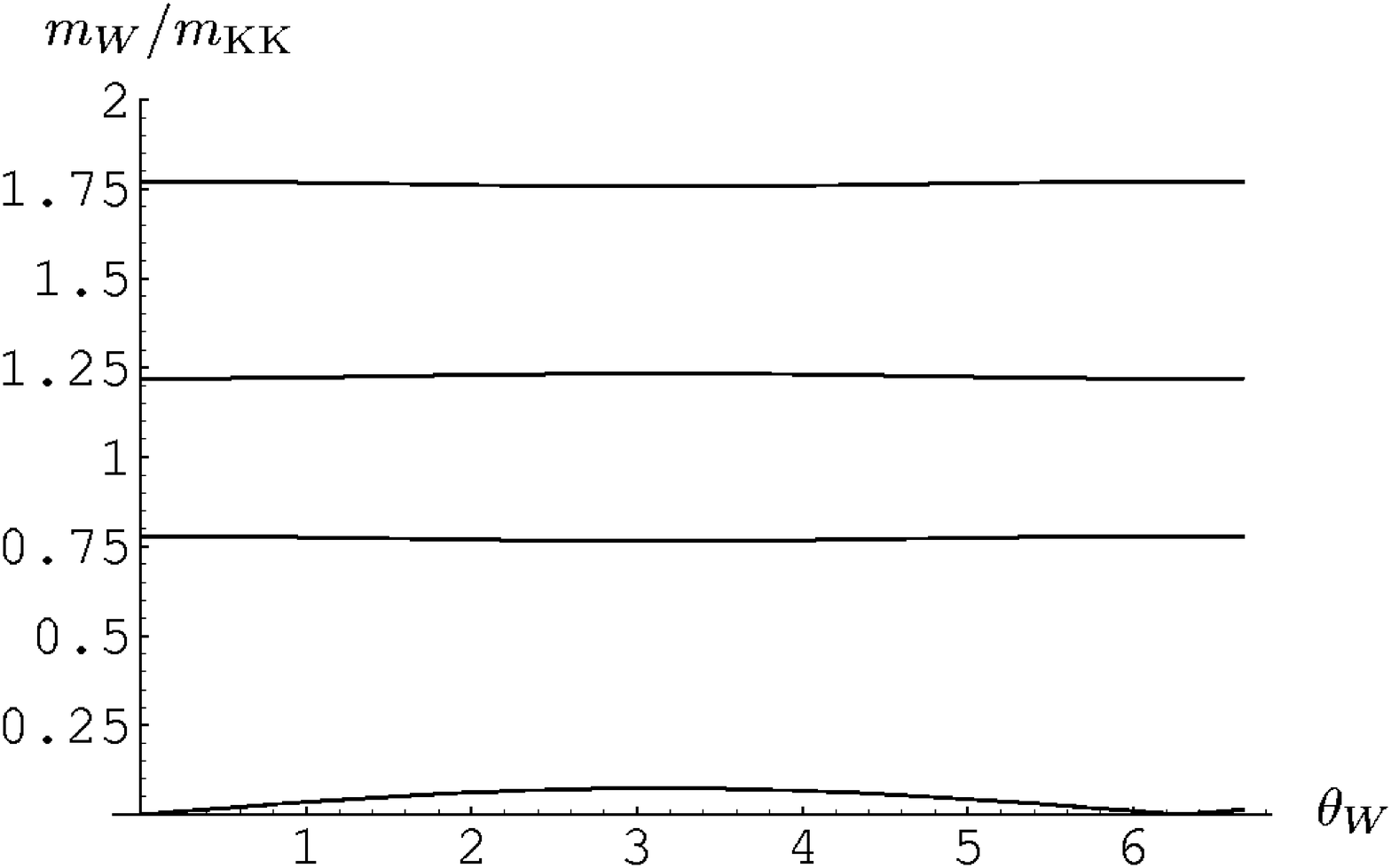}
\includegraphics[width=5.2cm]{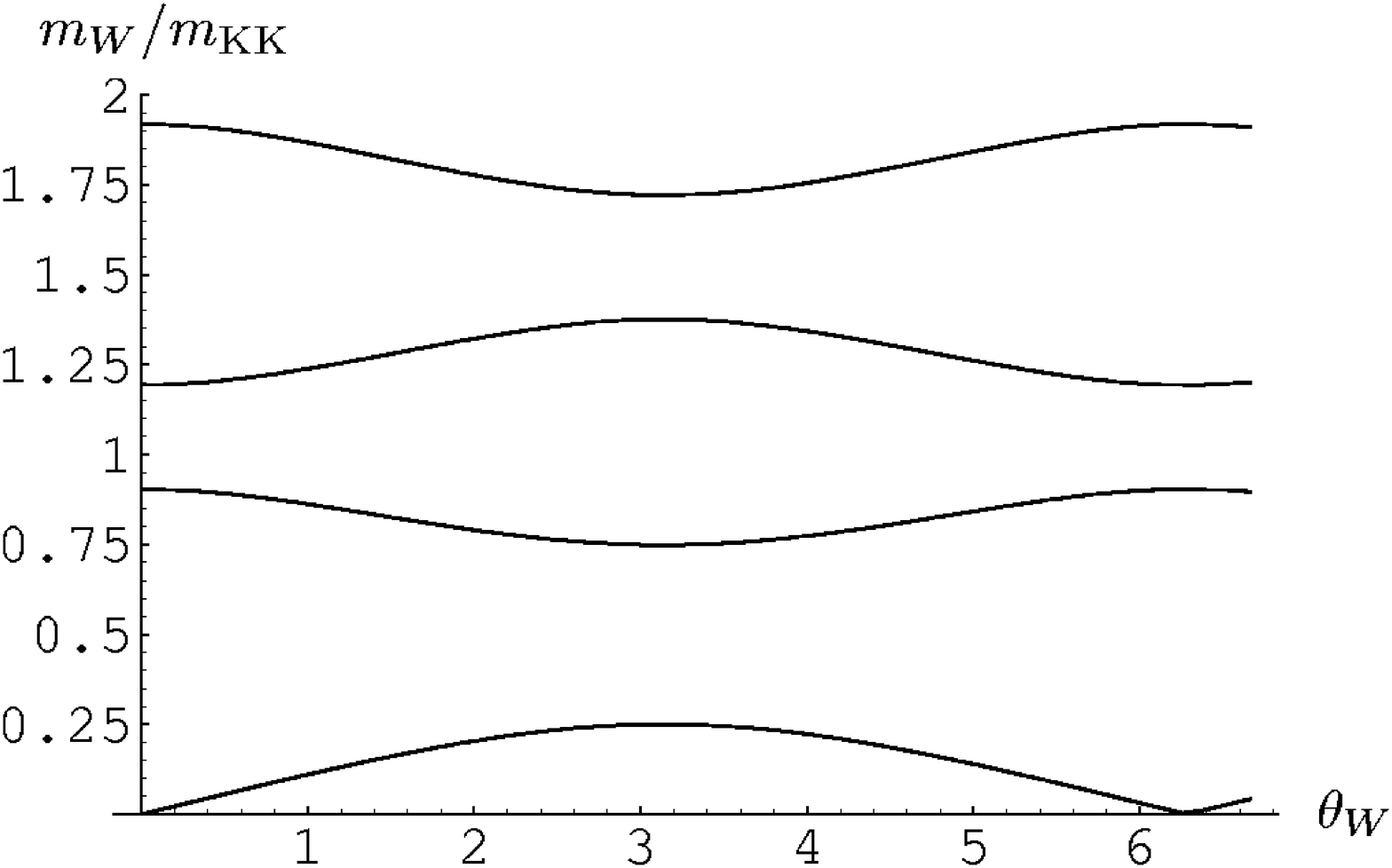}
\includegraphics[width=5.2cm]{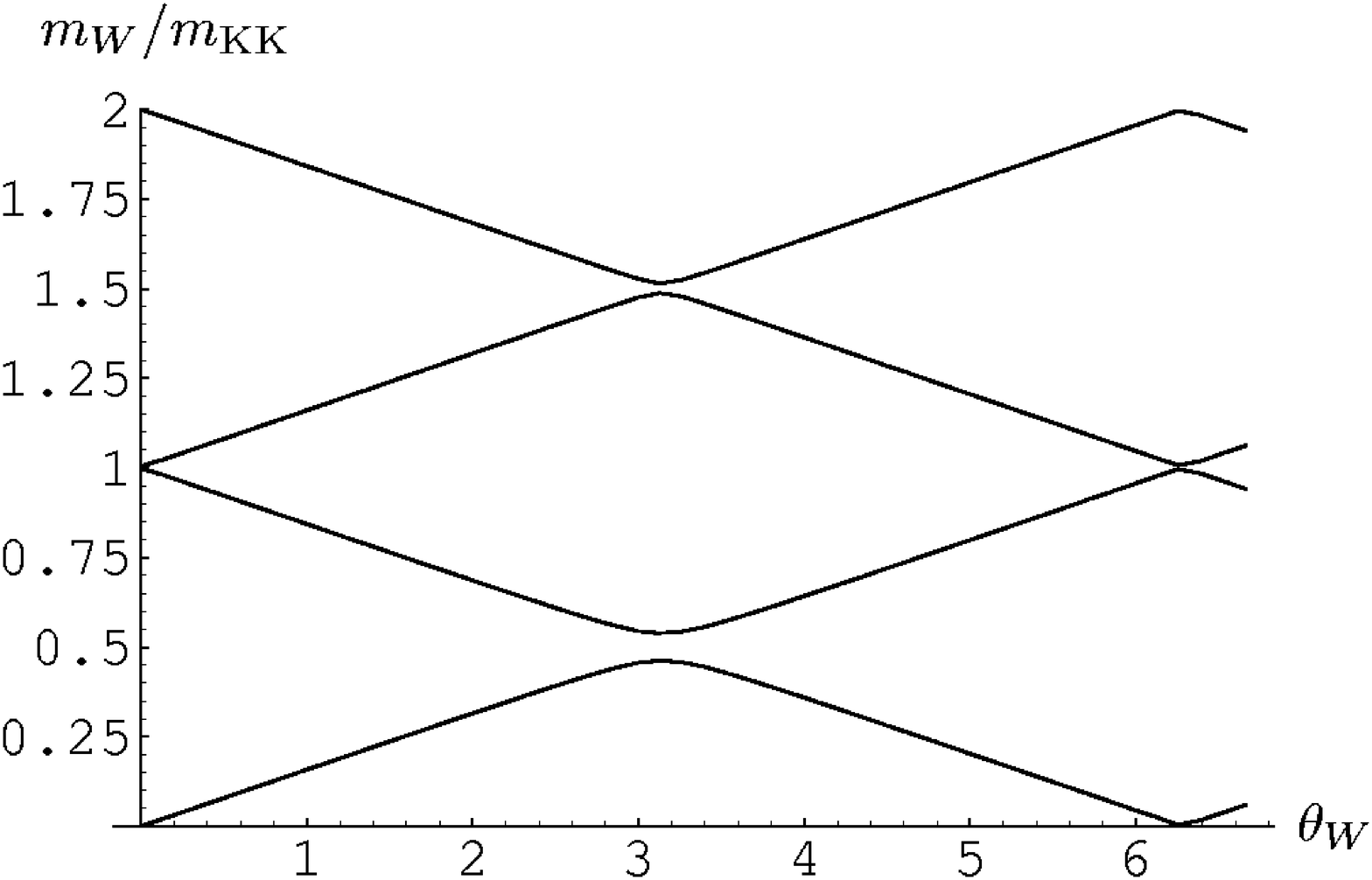}
\caption{The masses of $W$ boson and its KK excited states
are depicted in the unit of $m_{\rm KK}$ as functions of $\thw$
for $kR=12, 1.2$, and $0.12$ from the left to the right.
}
\label{Wmass_thw}
\end{figure}

As is evident from (\ref{mass_det}), all mass eigenvalues are 
periodic in $\thw$; $m_n(\theta_W + 2\pi) = m_n(\theta_W)$.
We remark that this behavior is in no contradiction to the 
behavior observed in flat space 
$m_n(\theta_W + 2\pi) = m_{n+\ell}(\theta_W)$ 
($\ell$: an integer).\footnote{It has been argued 
in Ref.\ \cite{HM} that $\ell$ is, in general, a non-vanishing integer.
It turns out that $\ell=0$ for every field in the curved space.
In either case the spectrum itself is periodic in $\theta_W$ so that
the argument in Ref.\ \cite{HM} remains valid.}
In order to understand the situation, let us see the mass spectrum 
for massive modes. 
By utilizing the asymptotic behavior of the Bessel 
functions~(\ref{asymp_bhv}),  the relation (\ref{mass_det}) becomes,
for $\lmd_n\gg 1$, 
\be
 \sin^2 \big\{ (\zp-1)\lmd_n \big\} \simeq \sin^2\hthw ~, 
\ee
which leads to the mass spectrum\footnote{The formula 
(\ref{hv_mn}) is valid independent of the value of $\alp$.}
\be
 m_n \simeq \abs{n+\frac{\thw}{2\pi}}m_{\rm KK} ~.
  \label{hv_mn}
\ee
In  flat spacetime,  all $\lambda_n$'s become large except
for the zero mode $\lambda_0=0$ so that the formula 
(\ref{hv_mn}) becomes exact for all $n$'s.
The behavior (\ref{hv_mn}) can be interpreted that 
 each mass eigenvalue shifts to the next KK level 
as $\thw\to\thw+2\pi$,  or equivalently that 
$m_n(\theta_W + 2\pi) = m_{n+1}(\theta_W)$.
In the curved space, however, this is incorrect.
Fig.~\ref{Wmass_thw} depicts the masses of $W$ boson and 
its KK modes as functions of $\thw$. 
It shows that each mass eigenvalue is periodic in $\thw$, 
the level-crossing never taking place. 
As the AdS curvature~$k$ becomes small, two adjacent lines come
closer to each other at $\theta_W=\pi$, attaching to each other
in the flat limit ($k \go 0$).  It can be said that the level 
crossing occurs in the flat space.   However, for $k>0$ two lines
never cross each other, and 
(\ref{hv_mn}) should be written as 
\be
 m_n \simeq \abs{n+\frac{1}{2}-\frac{\abs{\pi-\thw}}{2\pi}} m_{\rm KK} 
 \label{hv_mn2}
\ee
for $0\leq\thw \leq 2\pi$. 
In the flat limit, this amounts to relabeling the KK modes, 
but only (\ref{hv_mn2}) describes the correct $\thw$-dependence of 
a mass-eigenvalue  in the warped case ($k\neq 0$). 

\subsection{The quark-lepton mass spectrum}

The mass spectrum in the fermion sector depends on the bulk mass~$M$ 
through $\alp = (M/k) + \onehalf$ in (\ref{mass_det}), which can 
be used to reproduce the  mass spectrum of quarks and leptons.
This is possible only in the warped spacetime, since
 the mass spectra for all fields 
are independent of $M$ in the flat case.  

Before proceeding, we need to specify the fermion content in more
detail.  As a typical example we consider a model adopted
in Ref.\ \cite{Antoniadis1}, in which 
\beeq
\begin{pmatrix}
\nu_{e{\rm L}} &\tilde \nu_{e{\rm R}} \cr 
e_{\rm L}  & \tilde e_{\rm R}\cr 
\tilde e_{\rm L} & e_{\rm R}
\end{pmatrix}
~,~
\begin{pmatrix}
\tilde \nu_{e\rm L} &\nu_{e \rm R} 
\end{pmatrix}
~,~
\begin{pmatrix}
{d^c}_{{\rm L}}  &\tilde {d^c}_{{\rm R}}\cr
 {u^c}_{\rm L} & \tilde {u^c}_{\rm R} \cr 
 \tilde {u^c}_{\rm L} & {u^c}_{\rm R}
\end{pmatrix}
~,~
\begin{pmatrix}
\tilde {d^c}_{\rm L} &{d^c}_{\rm R} 
\end{pmatrix}
\label{Fcontent}
\eneq
are contained in the first generation.  Boundary conditions are
chosen such that only fields without tilde have zero modes
for $\theta_W =0$.  In our scheme the zero modes of 
$e$ and $u$ acquire nonvanishing masses  when $\theta_W \not= 0$. 

\begin{figure}[b,t,h]
\centering  \leavevmode
\includegraphics[width=8.cm]{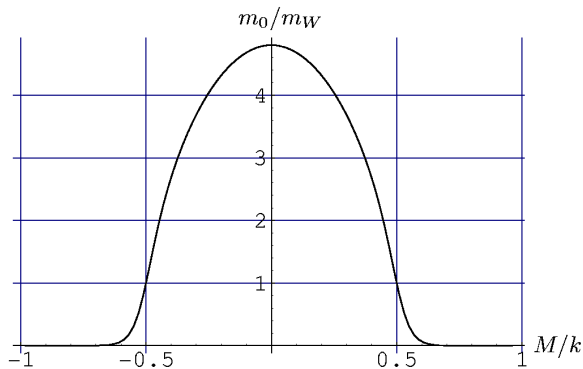}
\caption{The lightest mass eigenvalue~$m_0$ as a function of $M/k$
at $\theta_W=\onehalf \pi$. 
The vertical axis is in the unit of $m_W$, or 
the fermion mass value at $M/k=\pm \frac{1}{2}$.
$m_0/m_W$ has little dependence on $\theta_W$.}
\label{lmd0_c}
\end{figure}

The ratio of the lightest fermion mass $m_f(\alpha)$ to $m_W$ is almost
independent of the value of $\theta_W$.  
For $z_\pi \gg 1$, in particular, it follows from (\ref{ap_m0}) and
(\ref{ap_mW}) that
\be
 m_f(\alpha) \simeq 
 \brkt{ 
 \frac{\zp \alp(\alp-1) k\pi R}
 {2\sinh(\alp k\pi R)\sinh((\alp-1)k\pi R)}}^{\frac{1}{2}}
 m_W  ~.  \label{Fmass1}
\ee
Fig.~\ref{lmd0_c} shows the lightest mass~$m_0$ determined from (\ref{pms}) 
as a function of $M/k$ at $\theta_W=\onehalf \pi$. 
The fermion mass becomes the largest at $\alpha = \onehalf$ 
($M=0$).  Its value is  given by
$m_f(\onehalf) = 4.80 \, m_W = 386 \,$GeV for $kR=12$ and 
$\theta_W=\onehalf \pi$. 
Each $\lmd_n$ is an even function of $M$, as can be seen 
from (\ref{mass_det}) and (\ref{flip_F}). 
However, the corresponding mode functions are not invariant 
under $M \go -M$. (See Eq.(\ref{flip_md_fn}).)
$M/k$ is determined by the observed mass of quarks or leptons.
The determined values of $M/k$  are listed in Table~\ref{c_values}. 
Since $\lmd_n$ are even functions of $M$, a given value of $\lmd_n$ 
in general corresponds to two values of $M/k$. 
Only one of them can be consistent with the observation
as the other value leads to too large couplings to the
KK excited states of the $W$ boson, which will be 
detailed in the next section.

\begin{table}[b,t,h]
\begin{center}
\begin{tabular}{|c||c|c|c||c|c|c|}
\noalign{\kern 10pt}
\hline
\rule[-2mm]{0mm}{7mm} & $e$ & $\mu$ & $\tau$ & $u$ & $c$ & $t$ \\ \hline 
mass (GeV) & 5.11$\times 10^{-4}$ & 0.106 & 1.78 & 4$\times 10^{-3}$ & 1.3 &175 \\ \hline
$M/k$ & 0.865 & 0.715 & 0.633 & 0.81 & 0.64 &0.436 \\ \hline
\end{tabular}
\end{center}
\caption{The values of $M/k$ for leptons and quarks when $kR=12$ 
and $\theta_W=\onehalf \pi$.  The values have little dependence on $\theta_W$.}
\label{c_values}
\end{table}

Notice that the values of the dimensionless parameter $M/k$ for
all quarks and leptons fall in the range $(0.436 \sim 0.865)$.
Although there seems large hierarchy in the mass spectrum of
order $m_t/m_e \sim 10^5$, there is no such hierarchy in terms 
of the dimensionless parameter $M/k$ which is more natural
quantity in gauge theory in the RS spacetime. A similar 
role of the bulk mass in flat space has been pointed out in Ref.\cite{Arkani2}.
This might give an 
important hint in understanding the spectrum of quarks and leptons.

\subsection{The mass and self-couplings of the  Higgs boson}

The 4D Higgs field whose VEV breaks the gauge symmetry $SU(2) \times U(1)$
to $U(1)_{\rm EM}$ 
is identified with the zero-mode~$\vph_0$ in $A^7_z$ in the present scheme. 
At the classical level, the potential for $\vphi_0$ is flat and 
$\vphi_0$ is massless.   
The flat direction is parametrized by the Wilson line phase~$\thw$. 
At the quantum level the degeneracy is lifted, the effective potential
$V_\eff(\theta_W)$ becoming nontrivial.  From the global minimum of 
$V_\eff(\theta_W)$, the VEV of $\thw$ or $\vph_0$ is determined.
The Higgs field~$\vph_0$ acquires a nonvanishing finite mass. 
Although we do not explicitly calculate the effective potential
in this paper, we can draw many important conclusions from the mass
spectrum obtained in the preceeding sections.

The general argument given in Ref.\ \cite{HM} remains valid with 
minor changes resulting from the mass spectrum obtained in the 
preceeding sections. 
The one-loop effective potential~$V_{\rm eff}(\thw)$ in  flat spacetime 
has been evaluated well~\cite{Lim1,HHHK,HHK,HNT1}. 
$V_{\rm eff}(\thw)$ in the RS warped spacetime has been evaluated by Oda
and Weiler\cite{Oda1}.  At the one loop level
it depends only on the mass spectrum of  fields in the theory. 
Since the mass spectrum $\{ m_n \}$ in the warped space 
is almost the same for large $n$ as that in flat space
(see Eq.(\ref{hv_mn2})), 
the resultant  $V_{\rm eff}(\thw)$ takes a similar form
to that in the flat case\cite{HM}. 
It takes the form~\footnote{
As shown in section~\ref{pms}, $m_n(\theta_W +2\pi) = m_n(\theta_W)$
in the RS spacetime.  Accordingly the argument concerning 
 the spectrum~$\rho_n(\thw)$ in Eq.(22) of Ref.~\cite{HM} 
should be modified.  There
 $\rho_n(\thw+2\pi)=\rho_n(\thw)$, or $l$ is always zero. } 
\be
 V_{\rm eff}(\thw) = \frac{3}{128\pi^6}m_{\rm KK}^4 f(\thw), 
\ee
where $f(\thw)$ is a dimensionless periodic function of $\thw$ 
with a period $2\pi$. 
The explicit form of $f(\thw)$ depends on the matter content of the model, 
but its typical size is of order one 
in the minimal model or its minimal extension. 
When $V_{\rm eff}(\thw)$ has a global minimum at a nontrivial 
$\thw=\thw^{\rm min}$, 
 dynamical electroweak symmetry breaking takes place. 
 The global minimum is located typically 
 at $\thw^{\rm min}=(0.2\sim 1)\pi$ \cite{HNT2, YH6}. 
It is  possible to have a very small $\thw^{\rm min}\sim 0.01\pi$ 
by fine tuning of the matter content as shown in Ref.~\cite{HHKY}, 
but we will not consider such a case.  

The Higgs mass~$m_H$ is found by expanding $V_{\rm eff}(\thw)$ 
around $\thw^{\rm min}$. More generally one obtains 
\beeq
V_\eff =  \frac{3}{128\pi^6}m_{\rm KK}^4
\sum_n \myfrac{\phi^n}{n!} \bigg\{ 
\frac{2\pi^2 \alpha_W R (z_\pi^2 -1)}{k} \bigg\}^{n/2}
f^{(n)} (\thw^{\rm min}) 
\label{expandV}
\eneq
where $\alp_W\equiv g_4^2/4\pi$ and 
 $f^{(n)} (\thw) = d^n f(\thw)/d\thw^n$.
($g_4\equiv g_5/\sqrt{\pi R}$ is 
the 4D weak $SU(2)_L$ gauge coupling constant before the 
symmetry breaking takes place. See the next section.) 
The Higgs mass is evaluated from the $n=2$ term;
\be
 m_H^2 = f^{(2)}(\thw^{\rm min}) \, \frac{3\alp_W}{64\pi^4}
 \frac{R(\zp^2-1)}{k} \, m_{\rm KK}^4 ~~, 
 \label{Higgs1}
\ee
Using Eqs.(\ref{def_mKK}) and (\ref{ap_mW}), one finds 
\beqn
m_H \eql
  \brc{f^{(2)}(\thw^{\rm min}) \, \frac{3\alp_W}{64\pi^2}}^{\frac{1}{2}}
 \sqrt{kR} ~m_{\rm KK} \cr
\noalign{\kern 5pt}
\eql
\brc{f^{(2)}(\thw^{\rm min}) \, \frac{3\alp_W}{32\pi}}^{\frac{1}{2}}
 \frac{\pi kR}{2} 
\myfrac{m_W}{\sin \onehalf \thw^{\rm min}} ~~, 
\label{Higgs2} 
\eeqn
when $\zp\gg 1$. The important difference from the formula 
in flat space is the appearance of an enhancement factor
$\onehalf \pi kR \sim 18.8$.  In the model of \cite{HNT2}
it is found that $\big\{ f^{(2)}(\thw^{\rm min}) \big\}^{1/2} \sim 1.9$.
With this value inserted $m_H$ is predicted to be 286 Gev or 125 GeV for
$\thw^{\rm min} = 0.2\pi$ or $0.5\pi$, respectively. 
It is remarkable that the predicted value is exactly in the range
where experiments at LHC can explore.

The cubic and quartic coupling constants $\eta$ and $\lambda$ in the expansion
$V_\eff = \onehalf m_H^2 \phi^2 + \frac{1}{3}\eta \phi^3
+ \frac{1}{4}\lambda \phi^4 + \cdots$ are  given by 
\beqn
&&\hskip -1cm
\eta = \myfrac{3 \alpha_W^{3/2}m_W}{32 \pi^{1/2} \sin \onehalf \theta_W} 
f^{(3)} (\thw^{\rm min})  
\bigg( \frac{\pi kR}{2}  \bigg)^2 ~~, \cr
\noalign{\kern 5pt}
&&\hskip -1cm
\lambda = \myfrac{\alpha_W^2}{16} f^{(4)} (\thw^{\rm min}) \,
\bigg( \frac{\pi kR}{2}  \bigg)^2 ~~.
\label{Higgs3}
\eeqn
As in (\ref{Higgs2}), there appears an enhancement factor 
$(\onehalf \pi kR)^2$ for both $\eta$ and $\lambda$.  
In the standard model the relations 
$\eta = 3 \lambda v = 3\lambda m_W/\sqrt{\pi \alpha_W}$ and 
$\lambda = m_H^2/2v^2 = \pi m_H^2 \alpha_W/2m_W^2$ hold at the tree
level. In our scheme we find, instead, that
\beqn
&&\hskip -1cm
\eta \cdot \myfrac{\sqrt{\pi \alpha_W}}{3\lambda m_W}
= \myfrac{f^{(3)} (\thw^{\rm min})}{2 f^{(4)} (\thw^{\rm min})}
\myfrac{1}{\sin \onehalf \theta_W^{\rm min}} ~~, \cr
\noalign{\kern 10pt}
&&\hskip -1cm
\lambda \cdot \myfrac{2  m_W^2}{\pi  m_H^2 \alpha_W}
= \myfrac{4 f^{(4)} (\thw^{\rm min})}{3 f^{(2)} (\thw^{\rm min})}
\, \sin^2 \onehalf \thw^{\rm min} ~~.
\label{relation3}
\eeqn
The behavior of these couplings in flat space has been investigated
in Ref.\ \cite{Haba}.

\subsection{$\thw\to 0$ limit}

For the better understanding of the KK modes with respect to 
the broken  $SU(2)\times U(1)$ gauge symmetry, 
we  consider the $\thw\to 0$ limit. 
Take $(\psi_2,\psi_3)$ as an example. 
In this limit, the KK tower of $(\psi_2,\psi_3)$ 
splits into two KK towers of $\psi_2$ and $\psi_3$, 
and the $SU(2)\times U(1)$ gauge symmetry recovers. 
The mass spectrum determined from Eq.(\ref{mass_det})
splits into three cases.
\begin{enumerate}
 \item $\lmd_n\to 0$ \label{zrmd}
 \item $F_{\alp-1,\alp-1}(\lmd_n,\zp) \to 0$  \label{Z2evenL}
 \item $F_{\alp,\alp}(\lmd_n,\zp) \to 0$  \label{Z2oddL}
\end{enumerate}

The case~\ref{zrmd} corresponds to the zero modes ($n=0$) which
are contained in  the $Z_2$-even fields, \ie, 
$\psi_{{\rm L}2}$ and $\psi_{{\rm R}3}$. 
From (\ref{limit_F}), 
\bea
 F_{\alp,\alp}(\lmd_0,\zp) & \to & -\frac{2\sinh(\alp k\pi R)}{\pi\alp} ~, 
 \nonumber\\
 F_{\alp,\alp-1}(\lmd_0,\zp) & \to & \frac{2\zp^{-\alp}}{\pi\lmd_0} ~, 
 \nonumber\\
 F_{\alp-1,\alp}(\lmd_0,\zp) & \to & -\frac{2\zp^{\alp-1}}{\pi\lmd_0} ~,
\eea
so that
\be
 \lmd_0 = \brkt{\frac{\alp(\alp-1)}
 {\zp\sinh(\alp k\pi R)\sinh((\alp-1)k\pi R}}^{\frac{1}{2}}\frac{\thw}{2}
 +\cO(\thw^2), 
\ee
and the corresponding mode functions in (\ref{RH_md_fn}) 
and (\ref{LH_md_fn}) become
\bea
 \tl{f}^+_{2,0}(z) \eql \cO(\thw), \nonumber\\
 \tl{f}^+_{3,0}(z) \eql \sgn(\thw)
 \brkt{\frac{2k\alp}{\zp^{2\alp}-1}}^{\frac{1}{2}}
 z^{\alp-\frac{1}{2}}, \nonumber\\
 \tl{f}^-_{2,0}(z) \eql \brkt{
 \frac{2k(1-\alp)}{\zp^{2(1-\alp)}-1}}^{\frac{1}{2}}
 z^{\frac{1}{2}-\alp} = \tl{f}^-_{1,0}(z), \nonumber\\ 
 \tl{f}^-_{3,0}(z) \eql \cO(\thw).  \label{thw0lim1}
\eea
In the case~\ref{Z2evenL}, 
\be
 F_{\alp-1,\alp-1}(\lmd_n,\zp) = 
 \frac{\thw^2}{\pi^2\lmd_n^2\zp F_{\alp,\alp}(\lmd_n,\zp)}+\cO(\thw^3), 
\ee
while $\lmd_n$, $F_{\alp,\alp}(\lmd_n,\zp)$, $F_{\alp,\alp-1}(\lmd_n,\zp)$ and 
$F_{\alp-1,\alp}(\lmd_n,\zp)$ remain finite. 
Thus the mode functions become 
\bea
 \tl{f}^+_{2,n}(z) \eql \frac{\sqrt{k}\pi\lmd_n}{\sqrt{2}}
 \brc{\frac{Y_{\alp-1}^2(\lmd_n)}{Y_{\alp-1}^2(\lmd_n\zp)}-1}^{-\frac{1}{2}}
 z^{\frac{1}{2}}F_{\alp-1,\alp-1}(\lmd_n,z) = \tl{f}^+_{1,n}(z), \nonumber\\
 \tl{f}^+_{3,n}(z) \eql \cO(\thw), \nonumber\\
 \tl{f}^-_{2,n}(z) \eql \frac{\sqrt{k}\pi\lmd_n}{\sqrt{2}}
 \brc{\frac{Y_{\alp-1}^2(\lmd_n)}{Y_{\alp-1}^2(\lmd_n\zp)}-1}^{-\frac{1}{2}}
 z^{\frac{1}{2}}F_{\alp,\alp-1}(\lmd_n,z) = \tl{f}^-_{1,n}(z), \nonumber\\
 \tl{f}^-_{3,n}(z) \eql \cO(\thw). 
\eea
We observe that only $\psi_2$ components remain nonvanishing, which
form $SU(2)$-doublets with $\psi_1$ components in this limit. 
In the case~\ref{Z2oddL}, 
\be
 F_{\alp,\alp}(\lmd_n,\zp) = \frac{\thw^2}{\pi^2\lmd_n^2\zp 
 F_{\alp-1,\alp-1}(\lmd_n,\zp)}+\cO(\thw^2), 
\ee
while $\lmd_n$, $F_{\alp-1,\alp-1}(\lmd_n,\zp)$, $F_{\alp,\alp-1}(\lmd_n,\zp)$ 
and $F_{\alp-1,\alp}(\lmd_n,\zp)$ remain finite. 
Thus the mode functions become
\bea
 \tl{f}^+_{2,n}(z) \eql \cO(\thw), \nonumber\\
 \tl{f}^+_{3,n}(z) \eql p_{\alp,n}(\thw)\frac{\sqrt{k}\pi\lmd_n}{\sqrt{2}}
 \brc{\frac{Y_\alp^2(\lmd_n)}{Y_\alp^2(\lmd_n\zp)}-1}^{-\frac{1}{2}}
 z^{\frac{1}{2}}F_{\alp-1,\alp}(\lmd_n,z), \nonumber\\
 \tl{f}^-_{2,n}(z) \eql \cO(\thw), \nonumber\\
 \tl{f}^-_{3,n}(z) \eql p_{\alp,n}(\thw)\frac{\sqrt{k}\pi\lmd_n}{\sqrt{2}}
 \brc{\frac{Y_\alp^2(\lmd_n)}{Y_\alp^2(\lmd_n\zp)}-1}^{-\frac{1}{2}}
 z^{\frac{1}{2}}F_{\alp,\alp}(\lmd_n,z), 
\eea
where $p_{\alp,n}(\thw)$ is defined by (\ref{def_p}). 
Only $\psi_3$ components remain nonvanishing, forming 
$SU(2)$-singlets in this limit. 

The cases~\ref{Z2evenL} and \ref{Z2oddL} correspond to 
the non-zero modes. 
Their lightest modes have masses of order $m_{\rm KK}$ 
defined in (\ref{def_mKK}). 
The solutions~$\lmd_n$ of 
$F_{\alp,\alp}(\lmd_n,\zp)=0$ monotonically increase as $\abs{\alp}$ increases. 
Thus, if $\alp>\frac{1}{2}$ ($\alp<\frac{1}{2}$), 
the KK modes whose level number~$n$ is odd (even) belong to the case~\ref{Z2evenL},
and the modes with even (odd)~$n$ belong to the case~\ref{Z2oddL}. 
In the case of $\alp=\frac{1}{2}$, \ie, $M=0$, 
the modes in the cases~\ref{Z2evenL} and \ref{Z2oddL} are degenerate. 
In this case, all the Bessel functions appearing 
in mode functions reduce to trigonometric functions, and 
we can solve Eq.(\ref{mass_det}) analytically as 
\be
 m_n = \abs{n+\frac{1}{2}-\frac{\abs{\pi-\thw}}{2\pi}} \, m_{\rm KK} 
 \hskip 1cm
 (0\leq\thw \leq 2\pi)  
\ee
where $n=0,\pm 1,\pm 2,\cdots$. 
Here we have labeled the KK level number such that each KK 
mode is periodic in $\thw$. (See Sec.~\ref{pms}.) 

It is convenient to divide the KK modes into three types 
classified above.
In the case of the electron field ($\alp=1.37$), for example, 
we denote each KK mode as
\bea
 \tilde \psi_{{\rm L}2}(x,z) \eql \tl{f}^-_{2,0}(z)e_{{\rm L}0}(x)
 +\sum_{n=1}^\infty \tl{f}^-_{2,2n-1}(z)e_{{\rm L}n}(x)
 +\sum_{n=1}^\infty \tl{f}^-_{2,2n}(z)\tl{e}_{{\rm L}n}(x), \nonumber\\
 \tilde \psi_{{\rm L}3}(x,z) \eql \tl{f}^-_{3,0}(z)e_{{\rm L}0}(x)
 +\sum_{n=1}^\infty \tl{f}^-_{3,2n-1}(z)e_{{\rm L}n}(x) 
 +\sum_{n=1}^\infty \tl{f}^-_{3,2n}(z)\tl{e}_{{\rm L}n}(x), \nonumber\\
 \tilde \psi_{{\rm R}2}(x,z) \eql \tl{f}^+_{2,0}(z)e_{{\rm R}0}(x)
 +\sum_{n=1}^\infty \tl{f}^+_{2,2n-1}(z)\tl{e}_{{\rm R}n}(x)
 +\sum_{n=1}^\infty \tl{f}^+_{2,2n}(z)e_{{\rm R}n}(x), \nonumber\\
 \tilde \psi_{{\rm R}3}(x,z) \eql \tl{f}^+_{3,0}(z)e_{{\rm R}0}(x)
 +\sum_{n=1}^\infty \tl{f}^+_{3,2n-1}(z)\tl{e}_{{\rm R}n}(x)
 +\sum_{n=1}^\infty \tl{f}^+_{3,2n}(z)e_{{\rm R}n}(x) ~.
\label{expansion3} 
\eea
Similarly, the $W$ boson field ($\alp=1$) is expanded as
\bea
 \frac{1}{\sqrt{2}} \brkt{\tilde A^1_\mu+i\tilde A^2_\mu} 
 \eql \tl{h}^1_{A,0}(z)W_{\mu,0}(x)
 +\sum_{n=1}^\infty \tl{h}^1_{A,2n-1}(z)W_{\mu,n}(x)
 +\sum_{n=1}^\infty \tl{h}^1_{A,2n}(z)\tl{W}_{\mu,n}(x), \nonumber\\
 \frac{1}{\sqrt{2}} \brkt{\tilde A^4_\mu+i\tilde A^5_\mu} 
 \eql \tl{h}^4_{A,0}(z)W_{\mu,0}(x)
 +\sum_{n=1}^\infty \tl{h}^4_{A,2n-1}(z)W_{\mu,n}(x)
 +\sum_{n=1}^\infty \tl{h}^4_{A,2n}(z)\tl{W}_{\mu,n}(x). \nonumber\\
\label{expansion4} 
\eea
In the expansions (\ref{expansion3}) and (\ref{expansion4})
a tower of 4D fields with tilde does not have a zero-mode 
at $\theta_W=0$.

\section{Gauge couplings}

One of the startling consequences in the dynamical gauge-Higgs 
unification in the RS spacetime is the prediction of 
non-universality of weak interctions in the fermion sector.
With the wave functions of the gauge fields and quark-lepton fields
being established, one can unambiguiously determine gauge couplings
among them and their KK excited states from the observed
quark and lepton masses. 
The relevant terms in the action are
\bea
 I_{\rm gc} \eql \int d^5x \, \sqrt{-G} \, 
 g_5\bar{\psi}\Gm^M A_M\psi \cr
\noalign{\kern 5pt}
 \eql  \int d^4x \int_1^{z_\pi} \frac{dz}{k} \, 
g_5 \, \Big\{  \bar{\tl{\psi}}  \gamma^\mu \tilde A_\mu  \tl{\psi}
 +\cdots  \Big\} ~~.
 \label{gcAction1}
\eea
Inserting (\ref{expansion3}) and (\ref{expansion4}) into 
(\ref{gcAction1}), one obtains 
\be
 \cL^{(4)}_{\rm gc} = 
 \sum_n \frac{g_{(n)}}{\sqrt{2}}
 \brkt{\bar{e}_{{\rm L}0}\gm^\mu W_{\mu,n}\nu_{{\rm L}0}+\hc}+\cdots ~, 
\label{gcLagrangian}
\ee
where the ellipsis denotes terms involving the KK modes of the fermions 
or $\tl{W}_{\mu,n}$. 
Here the 4D gauge coupling constants are given by
\bea
g_{(0)}(\theta_W, \alpha)  \defa g_5\int_1^{\zp}\frac{dz}{k}
 \brkt{\tl{f}^-_{2,0}\tl{h}^1_{A,0}
 +\tl{f}^-_{3,0}\tl{h}^4_{A,0}}\tl{f}^-_{1,0} ~~,  \nonumber\\
\noalign{\kern 5pt}
g_{(n)}(\theta_W, \alpha) \defa g_5\int_1^{\zp}\frac{dz}{k}
 \brkt{\tl{f}^-_{2,0}\tl{h}^1_{A,2n-1}
 +\tl{f}^-_{3,0}\tl{h}^4_{A,2n-1}}\tl{f}^-_{1,0} \hskip .5cm (n\geq 1) ~, 
 \label{def_gc}
\eea 
which depend on $\theta_W$ and $\alpha= (M/k)+\onehalf$.

Consider the 4D weak gauge coupling~$g_{(0)}$. 
In the  $\thw\to 0$ limit the electroweak symmetry remains unbroken
so that $g_{(0)}$ must be universal for all quarks and leptons.
Indeed,  the mode functions of 4D gauge fields are constants
\be
 \tl{h}^1_{A,0}(z)\Big|_{\theta_W=0} 
 =\frac{1}{\sqrt{\pi R}} ~~,~~
 \tl{h}^4_{A,0}(z)\Big|_{\theta_W=0}  =0 ~~,  
 \label{h_thw0}
\ee
and  $\tl{f}^-_{2,0}(z)\big|_{\theta_W=0}$ coincides with
 $\tl{f}^-_{1,0}(z)$  (see Eq.\ (\ref{thw0lim1}))  so that
\be
 g_{(0)}(\thw=0, \alpha) = 
 g_5\int_1^{\zp}\frac{dz}{k}\frac{1}{\sqrt{\pi R}}
 \brkt{\tl{f}^-_{1,0}}^2 = \frac{g_5}{\sqrt{\pi R}}
 \equiv g_4  \label{def_g4}
\ee
where the normalization condition~(\ref{normalization2}) has been used.
$g_4$ is the 4D $SU(2)_L$ gauge coupling constant in the
unbroken theory. 
It is seen that $g_{(0)}$ is independent of $\alp$, 
or the fermion bulk mass~$M$, at $\theta_W=0$. 

When the electroweak symmetry breaking takes place so that 
$\theta_W \not= 0$, the overlap integral in (\ref{def_gc}) 
has nontrivial dependence on $\alp$. 
It leads to the violation of the universality in the couplings 
of the charged current to the $W$ boson.
In Table \ref{ge_values}, $g_{(0)}/g_4$ is tabulated for 
electrons for various values of $\theta_W$.  The deviation
of $g_{(0)}^{\rm electron}$ from $g_4$ remains very small.

\begin{table}[b,t]
\begin{center}
\begin{tabular}{|c|c|}
\noalign{\kern 10pt}
\hline
\rule[-2mm]{0mm}{7mm}  
$\theta_W$ & $g_{(0)}^{\rm electron}/g_4$  \\ \hline 
0 & 1  \\ \hline
$0.2 \pi$ & 1.00092  \\ \hline
$0.5 \pi$ & 1.00489  \\ \hline
$\pi$ & 1.00999  \\ \hline
\end{tabular}
\end{center}
\caption{$g_{(0)}/g_4$ for electrons when $kR=12$.}
\label{ge_values}
\end{table}

\begin{figure}[t]
\centering  \leavevmode
\includegraphics[width=8.cm]{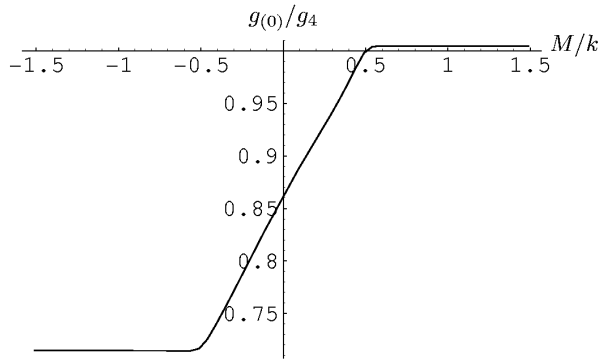}
\caption{The 4D gauge coupling~$g_{(0)}/g_4$ as a function of $M/k$
for $\theta_W=\onehalf \pi$ and $kR=12$.}
\label{g0-c}
\end{figure}

The dependence of $g_{(0)}/g_4$ on the fermion mass, or on $M/k$, 
is depicted in Fig.\ \ref{g0-c}.
It shows that $g_{(0)}$ is almost  constant for 
$\abs{M}/k >\frac{1}{2}$. 
This feature is understood from the profiles of the mode functions. 
For $M/k>\frac{1}{2}$ ($\alp>1$), 
\bea
 \tl{f}^-_{2,0}(z) &\simeq & \tl{f}^-_{1,0}(z) \simeq 
 \sqrt{2k(\alp-1)}z^{\frac{1}{2}-\alp}, \nonumber\\
 \tl{f}^-_{3,0}(z) &\simeq & \sqrt{\frac{k(\alp-1)}{2}}
 \frac{\sin\thw}{\zp^{2\alp}}z^{\frac{1}{2}+\alp}, 
 \label{pos_c}
\eea
and for $M/k<-\frac{1}{2}$ ($\alp<0$), 
\bea
 \tl{f}^-_{2,0}(z) &\simeq & \abs{\cos\hthw} \tl{f}^-_{1,0}(z) \simeq 
 \abs{\cos\hthw} \frac{\sqrt{2k(1-\alp)}}{\zp^{1-\alp}}z^{\frac{1}{2}-\alp}, 
 \nonumber\\
 \tl{f}^-_{3,0}(z) &\simeq & p_0\sin\hthw \tl{f}^-_{1,0}(z) 
 \simeq p_0\sin\hthw \frac{\sqrt{2k(1-\alp)}}{\zp^{1-\alp}}z^{\frac{1}{2}-\alp}, 
 \label{neg_c}
\eea
where $p_0\equiv\sgn(\cos\frac{1}{2}\thw)$. 
Here we have made use of the fact that the  
eigenvalue~$\lmd_0$ is exponentially small 
for $\abs{M}/k>\frac{1}{2}$ (see Fig.~\ref{lmd0_c}) and Eq.(\ref{limit_F}). 
For $M/k>\frac{1}{2}$, we can see from Eq.(\ref{pos_c}) that 
the first terms in Eq.(\ref{def_gc}) dominate 
and $\tl{f}^-_{1,0}(z)\tl{f}^-_{2,0}(z)/k$ plays a  role similar to that of
the delta function~$\dlt(z-1)$ since it is strongly localized around $z=1$. 
For $M/k<-\frac{1}{2}$,  all fermion mode functions 
in Eq.(\ref{def_gc}) are dominant around $z=\zp$,  thus  picking up 
the value of $\tl{h}^{1,4}_{A,0}(z)$ in the vicinity of $z=\zp$. 
As a consequence the gauge coupling~$g_{(0)}$ is almost independent of $\alp$ 
for $\abs{M}/k>\frac{1}{2}$. 
The asymptotic values of $g_{(0)}$ in this region is evaluated by 
the following behavior of the mode functions for the gauge fields. 
\bea
 \tl{h}^1_{A,0}(1) &\simeq & \frac{1}{\sqrt{\pi R}}, \nonumber\\
 \frac{\tl{h}^1_{A,0}(\zp)}{\tl{h}^1_{A,0}(1)} &\simeq & \cos^2\hthw, \;\;\;\;\;
 \frac{\tl{h}^4_{A,0}(\zp)}{\tl{h}^1_{A,0}(1)} \simeq \sin\hthw\cos\hthw. 
\eea
Here we have used Eq.(\ref{ap_mW}). 
Using these relations, $g_{(0)}$ is evaluated to be 
\bea
 g_{(0)}\brkt{\frac{M}{k}>\frac{1}{2}} & \simeq & g_4 ~~, 
 \nonumber\\
 g_{(0)}\brkt{\frac{M}{k}<-\frac{1}{2}} & \simeq & g_4
 \Big| \cos\hthw  \Big|~~. 
 \label{asymp_g0}
\eea
There arise small corrections to the asymptotic values above
due to the extended nature of the fermion mode functions.

\ignore{
}

In view of stringent experimental constraints on the gauge 
coupling universality, we need to examine precisely how 
much deviation from the universality results.
For quarks and leptons with the values of $M/k$ in Table~\ref{c_values}, 
the couplings to the $W$ boson $g_{(0)}$ are evaluated from
(\ref{def_gc}).  The quantity of physical interest is the degree of
the violation of the universality.  Therefore, for each quark or 
lepton, $g_{(0)}/g_{(0)}^{\rm electron} -1$ is tabulated in
Table~\ref{g0_values}. 
It is seen that the violation of  the universality in the weak
gauge coupling is within the current experimental bounds. 
Violation of the universality becomes larger for heavy fermions. 

\begin{table}[b,t]
\begin{center}
\begin{tabular}{|c||c|c|c|} 
\noalign{\kern 15pt}
\hline
$\theta_W$ & $\mu$(muon) & $\tau$(tau) & $t$(top)  \\ \hline 
~$0.2\pi$~ & $- 1.74 \times 10^{-9}$ 
   & $-4.69 \times 10^{-7}$ & $-4.3 \times 10^{-3}$ \\ \hline
$0.5\pi$ & $- 9.26 \times 10^{-9}$ 
   & $-2.50 \times 10^{-6}$ & $-2.2 \times 10^{-2}$ \\ \hline
$0.8\pi$ & $- 1.70 \times 10^{-8}$ 
   & $- 4.60 \times 10^{-6}$ & $-4.0 \times 10^{-2}$ \\ \hline
\end{tabular}
\end{center}
\caption{Non-universality of weak interactions. 
The deviation of $g_{(0)}^{\rm f}/g_{(0)}^{\rm electron}$ from 1 for 
f$\, = \mu, \tau, t$ is listed. ($kR=12$.)
}
\label{g0_values}
\end{table}

\begin{figure}[t]
\centering  \leavevmode
\includegraphics[width=10.cm]{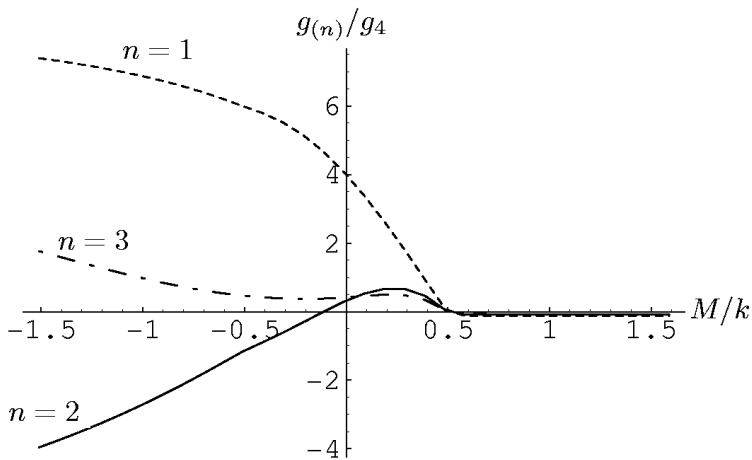}
\caption{The gauge couplings to the $n$-th KK excited states of 
$W$, $g_{(n)}$ ($n=1,2,3$) as functions of $M/k$ 
at $\theta_W=\onehalf\pi$. }
\label{gn-c}
\end{figure}

Next, we  consider the couplings to the KK excited states of
$W$ boson, $g_{(n)}$.  In Fig.~\ref{gn-c}, $g_{(n)}/g_4$ ($n=1,2,3$) 
are shown as functions of $M/k$ at $\theta_W=\onehalf \pi$. 
These quantities at $\theta_W=0$ have been evaluated by 
Gherghetta and Pomarol\cite{GP}.  Qualitative behavior does not 
change for $\theta_W \not= 0$.

Since  for $M/k =\frac{1}{2}$ ($\alpha=1$)
\bea
 z^{\frac{1}{2}}\tl{f}^-_{1,n}(z) \eql \frac{1}{\sqrt{\pi R}} ~, \cr
\noalign{\kern 5pt}
 z^{\frac{1}{2}}\tl{f}^-_{2,n}(z) \eql \tl{h}^1_{A,n}(z) ~, \cr
\noalign{\kern 10pt}
 z^{\frac{1}{2}}\tl{f}^-_{3,n}(z) \eql \tl{h}^4_{A,n}(z) ~, 
\eea
it follows with  the normalization condition in (\ref{normalization1})
that the 4D gauge couplings satisfy
\be
 g_{(0)}(\theta_W, 1) = g_4 ~~,~~
 g_{(n)}(\theta_W, 1) = 0 ~~.  \label{gn_hfc}
\ee
In the $\thw \go 0$ limit, this feature has been 
explained in Ref.\ \cite{GP} as a result of 
the accidental conformal symmetry 
or the translational invariance along the fifth direction. 
We have seen that the relation (\ref{gn_hfc})  holds even in the case 
 $\thw\neq 0$ 
where such accidental symmetry no longer exists. 

As mentioned in the derivation of (\ref{asymp_g0}), 
the fermion mode functions are strongly localized at the boundary 
$z=1$ when $M/k>\frac{1}{2}$. 
Due to this property, $g_{(n)}$ ($n\geq 1$) are almost independent 
of $\alp$ for $M/k>\frac{1}{2}$ just like $g_{(0)}$. 
On the other hand, they still have nontrivial $\alp$-dependences 
around $M/k\simeq -1$ in Fig.~\ref{gn-c}. 
This is because $\tilde h_{A,n}^{1,4}(z)$ ($n\geq 1$) oscillate around $z=\zp$ 
where the fermion mode functions are dominant. 
It has been argued in Ref.\ \cite{Chang} that brane fermions have
universal coupling $\abs{g_{(n)}}/g_4=\sqrt{2k \pi R}$.
We have numerically confirmed that 
$\abs{g_{(n)}}/g_4$ ($n=1,2,3$) in Fig.\ \ref{gn-c} approach
the asymptotic value $\sqrt{2k \pi R} \simeq 8.68$ as $M/k \go -\infty$.

\ignore{
In the limit of $\abs{M}/k\to -\infty$, 
both $\tl{f}^-_{1,0}(z)\tl{f}^-_{2,0}(z)/k$ and $\tl{f}^-_{1,0}(z)\tl{f}^-_{3,0}(z)$ 
are proportional to $\dlt(z-\zp)$, and $g_{(n)}$ will approach 
to some finite values.\footnote{
This limit corresponds to the case that the fermion zero-modes live 
on the TeV brane at $z=\zp$, and the limit values~$\abs{g_{(n)}}$ ($n\geq 1$) 
are universal, \ie $\abs{g_{(n)}}/g_4=\sqrt{2\pi kR}\simeq 8.4$ \cite{Chang}. }
}

We note  that a given value of $\lmd_n$ determines only the absolute
value of $M/k$ while its sign remains undetermined. 
(See Fig.~\ref{lmd0_c}.)  
The values of $M/k$ in Table~\ref{c_values} are 
the only  values consistent with experiments, as fermions with
the values of the opposite sign have too large $g_{(n)}$ leading to
contradiction to  precision measurements as discussed by
Chang et al\cite{Chang}.
The couplings $g_{(1)}/g_4$, $g_{(2)}/g_4$, and $g_{(3)}/g_4$ 
approach  $-0.13$, $-0.090$, and $-0.073$ for $M/k > 0.6$, respectively.

\section{Yukawa couplings}

The Yukawa couplings in four dimensions originate from 
five-dimensional gauge interactions in the dynamical gauge-Higgs unification 
scheme. From the 5D interaction 
\bea
 \cL_{\rm yukawa} \eql \sqrt{-G}g_5\bar{\psi}\Gm^4 A_z\psi \cr
 \eql  g_5(\bar{\tl{\psi}}_1,\bar{\tl{\psi}}_2,\bar{\tl{\psi}}_3)\gm_5 
 \frac{1}{2}\brkt{\begin{array}{ccc} & & \\ & & -i\tl{A}^7_z \\ 
 & i\tl{A}^7_z & \end{array}}
 \brkt{\begin{array}{c} \tl{\psi}_1 \\ \tl{\psi}_2 \\ \tl{\psi}_3 \end{array}}
\eea
the  4D coupling
\be
 \cL^{(4)}_{\rm yukawa} = -iy_e \vph_0\bar{e}_{{\rm L}0}e_{{\rm R}0}+\hc
\ee
emerges.  
Here the Yukawa coupling constant~$y_e$ is given by 
\be
 y_e \equiv \frac{g_5}{2}\int_1^{\zp}\dr z\;\tl{h}^7_{\vph,0}
 \brkt{\tl{f}^-_{2,0}\tl{f}^+_{3,0}-\tl{f}^-_{3,0}\tl{f}^+_{2,0}},  
 \label{Yukawa1}
\ee
where 
\be
 \tl{h}^7_{\vph,0}(z) = \sqrt{\frac{2}{k(\zp^2-1)}} ~ z. 
\ee

\begin{figure}[t]
\centering \leavevmode
\includegraphics[width=10.cm]{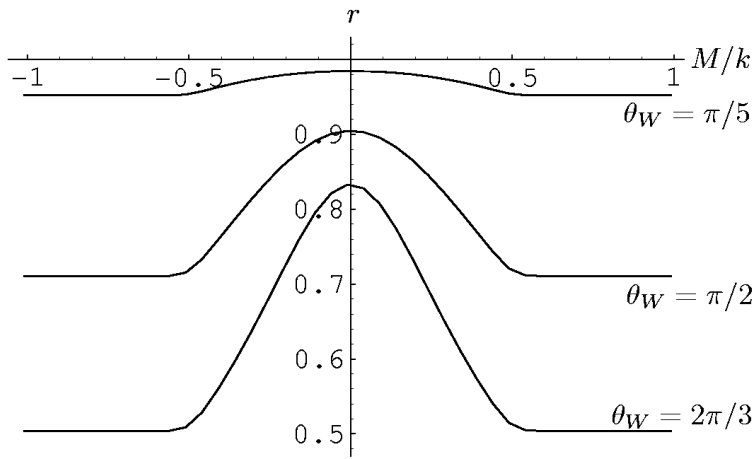}
\caption{The ratio~$r$ defined in Eq.(\ref{def_r}) as a function of $M/k$. 
$\theta_W=  \frac{1}{5}\pi, \onehalf\pi, \frac{2}{3}\pi$. 
Significant reduction of the Yukawa couplings is seen. The asymptotic 
value of $r$ for $|M/k| \gg \onehalf$ is $|\cos \onehalf\theta_W |$.}
\label{r-c}
\end{figure}

In accordance with the standard model, we define 
the ``Higgs VEV''~$v$ by
\be
 v \equiv \frac{2m_W}{g_4} \sim 246 \, {\rm GeV}~.
\ee
$g_4$ is  defined in Eq.(\ref{def_g4}). 
In the standard model, the fermion mass, say, the electron mass~$m_e$ is 
related to  $v$ and $y_e$ by $m_e = \abs{y_e} v$. 
Hence we define 
\be
 r \equiv \frac{2\abs{y_e}m_W}{g_4 m_e} ~,  \label{def_r}
\ee
which equals one for all fermions in the standard model. 
In the present scheme, this ratio~$r$ is not a constant 
but has a distinct value  for each fermion. 
In Fig.~\ref{r-c}, we plot $r$ as a function of $M/k$. 
Note that $r$ is an even function of $M/k$ as $y_e$ is even under 
$M\exch -M$.  (See Eq.(\ref{flip_md_fn}).) 
For $M=0$, all the Bessel functions appearing in the mode functions reduce 
to trigonometric functions, and we can analytically calculate $r$ as 
\be
 r(0) \simeq \frac{2}{\thw} \sin\hthw, 
\ee
for $-\pi\leq\thw \leq \pi$. 
The approximate expression of $m_W$~(\ref{ap_mW}) has been made use of. 
For $\abs{M}/k>\frac{1}{2}$, $r$ is almost independent of $M/k$.
From Eqs.(\ref{pos_c}), (\ref{neg_c}) and (\ref{flip_md_fn}), 
the asymptotic constant value is evaluated to be
\be
 r\brkt{\frac{\abs{M}}{k}>\frac{1}{2}} \simeq \abs{\cos\hthw} ~. 
 \label{YukawaRatio}
\ee

This is a stringent result.  As $\theta_W$ approaches $\pi$,
the asymptotic value of $r$, namely the Yukawa coupling,  
vanishes.  The significant reduction
in the Yukawa couplings result for all quarks and leptons for
$|\theta_W| > 0.3\pi$.  The measurement of the Yukawa couplings
definitely sheds light on the origin of the Higgs field.

\section{Conclusion and discussions}

In the present paper we have shown that many intriguing 
phenomenological consequences follow in the quark-lepton sector
in the scheme of the dynamical gauge-Higgs unification in the 
RS warped spacetime. 
The 4D Higgs field is identified with the extra-dimensional component 
of the gauge potentials, representing fluctuations of the 
Wilson line phase $\theta_W$. Dynamical electroweak symmetry 
breaking takes place when the phase $\theta_W$ takes a 
nontrivial value by quantum effects.  

Another important quantity for discussing fermion phenomenology
is the bulk mass $M$.  The dimensionless parameter
$c\equiv M/k$, where $-k^2$ is the AdS curvature in the bulk, 
becomes crucial for controling the wave functions of quarks
and leptons.  When $\theta_W=0$, quarks and leptons remain
massless irrespective of the value of $c$ although their
wave functions depend on $c$.  
As $\theta_W \not= 0$ and the dynamical electroweak 
symmetry breaking is induced, quarks and leptons acquire
nonvanishing masses, $m_f$, which depend on both $\theta_W$
and $c$.  Remarkably $m_f/m_W$ has little dependence on
$\theta_W$ so that the parameter $c$ is unambiguiously
determined by the observed quark and lepton masses.  
The values of $c$ are 0.865 and 0.436 for electrons and
top quarks, respectively.  Although there is huge
hierarchy in the fermion masses, there does not appear 
such hierarchic structure in $c$, the more natural
entity in the RS spacetime.

With the value of $c$ fixed for each fermion, one can
make important predictions for the gauge and Yukawa 
couplings of quarks and leptons.  First, non-universality
in the weak interactions results for the couplings of quarks
and leptons to the $W$ boson.  The magnitude of the 
non-universality is, however, very tiny.  We have found
that the violation of the universality for 
$\theta_W=\onehalf \pi$ is of magnitudes of
$9.26 \times 10^{-9}$, $2.50 \times 10^{-6}$, and 
$2.2 \times 10^{-2}$ for $\mu$-$e$, $\tau$-$e$, and 
$t$-$e$, respectively.  These numbers are well within the 
current experimental limits.  Improvement of experiments
is necessary to confirm the non-universality of the weak 
interactions.

Secondly, the Yukawa coupling of quarks and leptons to
the 4D Higgs field suffers from large corrections.
Compared with the Yukawa coupling in the standard model,
the Yukawa coupling in the dynamical gauge-Higgs unification
is suppressed by a factor $\sim |\cos \onehalf \theta_W|$.
This effect is observable at LHC, should the direct Yukawa
coupling be measured.

The non-universality of the weak interactions and the 
reduction of the Yukawa coupling are the two major predictions
we obtained in the present paper.  
As mentioned in the introduction, the scenario of the dynamical 
gauge-Higgs unification is by no means complete in the current form, however. 
The non-universality of the weak interactions is expected for 
neutral currents as well. To tackle this problem definitively, however,
one first needs to improve the model so as to have the correct value for
the Weinberg angle.  
The improvement along this direction is important considering that  constraints on 
the non-universality of the couplings to the $Z$ boson are much severer 
than those for  the $W$ boson discussed in this paper\cite{Agashe3}. 
One model with the correct value for the Weinberg angle has been provided
in Ref.\ \cite{Agashe2}, for which it is desired to perform analysis
outlined in the present paper. 
Secondly, right-handed components of fermions are dominantly localized
near the TeV brane at $z=z_\pi$ so that they are expected to have too large 
couplings to the Kaluza-Klein excited states of neutral gauge bosons, 
which may contradicts with the current precision measurements.
Additional structure might be required to obtain a realistic model.
Thirdly, neutrinos and down-type quarks, in the 
present minimal model, remain massless.  
The origin of masses of those fermions need to be clarified.  
Fourthly, in the pure gauge interactions 
it seems very difficult to accommodate CP violation 
as stressed by Frere\cite{Frere}. 
The third and fourth problems may suggest the existence of 
a fundamental scalar field. 
Further, implications to the $S$, $T$ parameters and the unitarity 
need to be investigated.   
We shall come back to these issues in the near future.

\vskip .8cm

\leftline{\bf Acknowledgments}
We would like to thank S.\ Kanemura for many enlightening comments.
This work was supported in part by  Scientific Grants from the Ministry of 
Education and Science, Grant No.\ 17540257, Grant No.\ 17043007, 
 Grant No.\ 13135215,  and Grant No.\ 15340078 (Y.H), 
 and by JSPS fellowship No.\ 0509241 (Y.S.)



\appendix

\section{Useful formulae for Bessel functions} 
 \label{usfl_fml}

In this appendix  we  collect  useful formulae for the Bessel functions
frequently used in the text. 
$J_\alp(z)$ and $Y_\alp(z)$ denote the Bessel functions of 
the first and second kinds, respectively. 
For $z\ll 1$, 
\be
 J_\alp(z) = \frac{1}{\Gm(\alp+1)}\brkt{\frac{z}{2}}^\alp\brc{1+\cO(z^2)} ~.
 \label{J_exp}
\ee
$Y_\alp(z)$ is defined as 
\be
Y_\alp(z) \equiv 
\begin{cases}
\myfrac{1}{\sin\pi\alp}\brc{\cos\pi\alp\cdot J_\alp(z)
 -J_{-\alp}(z)} &\hbox{for } \alpha \not= \hbox{ an integer,}\cr
\noalign{\kern 10pt}
 \myfrac{1}{\pi}\sbk{\myfrac{\der J_\alp(z)}{\der\alp}
 -(-1)^n\myfrac{\der J_{-\alp}(z)}{\der\alp}}_{\alp=n}
   &\hbox{for } \alpha=n= \hbox{ an integer.}
\end{cases}
\ee
Their behavior for  $\abs{z}\gg 1$ is given by
\bea
 J_\alp(z) & \sim & \sqrt{\frac{2}{\pi z}}\cos\brkt{z-\frac{(2\alp+1)\pi}{4}}, 
 \nonumber\\
 Y_\alp(z) & \sim & \sqrt{\frac{2}{\pi z}}\sin\brkt{z-\frac{(2\alp+1)\pi}{4}}. 
 \label{asymp_bhv}
\eea
These Bessel functions~$Z_\alp(z)=J_\alp(z)$ or $Y_\alp(z)$ 
satisfy 
\bea
 Z_{\alp-1}(z)+Z_{\alp+1}(z) \eql \frac{2\alp}{z}Z_\alp(z), \nonumber\\
 \frac{dZ_\alp(z)}{dz} = \frac{\alp}{z}Z_\alp(z)-Z_{\alp+1}(z) 
 \eql Z_{\alp-1}(z)-\frac{\alp}{z}Z_\alp(z), \nonumber\\
 J_\alp(z)Y_{\alp-1}(z)-Y_\alp(z)J_{\alp-1}(z) \eql \frac{2}{\pi z}. 
 \label{fml_JY}
\eea
The following integral formula is useful for  determining
 normalization factors of mode functions. 
\bea
 \int^z\dr z\;zZ_\alp(\lmd z)\tl{Z}_\alp(\lmd z) \eql 
 \frac{z^2}{4}\brc{2Z_\alp(\lmd z)\tl{Z}_\alp(\lmd z)
 -Z_{\alp-1}(\lmd z)\tl{Z}_{\alp+1}(\lmd z)
 -Z_{\alp+1}(\lmd z)\tl{Z}_{\alp-1}(\lmd z)}, \nonumber\\
\eea
where $Z_\alp(z)$, $\tl{Z}_\alp(z)$ are linear combinations of 
$J_\alp(z)$ and $Y_\alp(z)$. 
Besides the Bessel functions, the following formula is also useful. 
\be
 \Gm(\alp)\Gm(1-\alp) = \frac{\pi}{\sin\pi\alp} ~~.
\ee

\section{Definition of various functions} \label{fml}

We  define 
\be
 F_{\alp,\bt}(\lmd,z) \equiv Y_\bt(\lmd)J_\alp(\lmd z)-J_\bt(\lmd)Y_\alp(\lmd z) ~. 
 \label{def_F}
\ee
With the aid of the third equation in (\ref{fml_JY}), 
it satisfies the relation, 
\be
 F_{\alp-1,\alp}(\lmd,z)F_{\alp,\alp-1}(\lmd,z)
 = F_{\alp-1,\alp-1}(\lmd,z)F_{\alp,\alp}(\lmd,z)-\frac{4}{\pi^2\lmd^2 z}. 
 \label{rel_for_F}
\ee
For $\lmd , \lambda z \ll 1$, 
\bea
 F_{\alp,\alp}(\lmd,z) & \to & -\frac{z^\alp-z^{-\alp}}{\pi\alp} ~, \nonumber\\
 F_{\alp,\alp-1}(\lmd,z) & \to & \frac{2z^{-\alp}}{\pi\lmd} 
  - \frac{\lambda z^\alpha}{2\pi \alpha (\alpha - 1)} ~, \nonumber\\
 F_{\alp-1,\alp}(\lmd,z) & \to & -\frac{2z^{\alp-1}}{\pi\lmd}
   + \frac{\lambda z^{-\alpha +1}}{2\pi \alpha (\alpha - 1)}  ~. 
 \label{limit_F}
\eea

From these functions, the coefficients in the normalized mode functions 
are written as 
\bea
 C^{\rm d}_{\alp,n}(\thw) \defa 
 \frac{\sqrt{2k}}{\zp}\brc{
 \frac{F_{\alp-1,\alp-1}^2}{\sin^2 \onehalf\thw}
 +\frac{F_{\alp,\alp-1}^2}{\cos^2 \onehalf\thw}
 -\frac{\pi^2\lmd_n^2}{\sin^2\thw}F^2_{\alp-1,\alp-1}F^2_{\alp,\alp-1}
 -\frac{4}{\pi^2\lmd_n^2\zp^2}}^{-\frac{1}{2}}, \cr
\noalign{\kern 10pt}
 C^{\rm s}_{\alp,n}(\thw) \defa 
 -\cot\hthw\cdot\frac{F_{\alp-1,\alp-1}}{F_{\alp-1,\alp}}
 \cdot C^{\rm d}_{\alp,n}(\thw) 
 = \tan\hthw\cdot\frac{F_{\alp,\alp-1}}{F_{\alp,\alp}}
 \cdot C^{\rm d}_{\alp,n}(\thw), 
 \label{def_F_Cs}
\eea
where $\lmd_n$ is a solution of Eq.(\ref{mass_det}), and 
the arguments of all $F_{\alp,\bt}$ in the definition 
of $C^{\rm d,s}_{\alp,n}$ are $(\lmd_n,\zp)$. 
In the second equation, we have used Eqs.(\ref{mass_det}) and (\ref{mass_det2}). 
Using the sign factors~$p_{\alp,n}(\thw)=\pm 1$ defined by 
\be
 p_{\alp,n}(\thw) \equiv \sgn\brkt{-\cot\hthw\cdot
 \frac{F_{\alp-1,\alp-1}(\lmd_n,\zp)}{F_{\alp-1,\alp}(\lmd_n,\zp)}} 
 =\sgn\brkt{\tan\hthw\cdot
 \frac{F_{\alp,\alp-1}(\lmd_n,\zp)}{F_{\alp,\alp}(\lmd_n,\zp)}}, 
 \label{def_p}
\ee 
$C^{\rm s}_{\alp,n}(\thw)$ can be rewritten as 
\be
 C^{\rm s}_{\alp,n}(\thw) = 
 p_{\alp,n}(\thw)\frac{\sqrt{2k}}{\zp}\brc{
 \frac{F_{\alp,\alp}^2}{\sin^2\onehalf\thw}
 +\frac{F_{\alp-1,\alp}^2}{\cos^2 \onehalf\thw}
 -\frac{\pi^2\lmd_n^2}{\sin^2\thw}F^2_{\alp,\alp}F^2_{\alp-1,\alp}
 -\frac{4}{\pi^2\lmd_n^2\zp^2}}^{-\frac{1}{2}}, 
 \label{def_Cs2}
\ee
where the arguments of $F_{\alp,\bt}$ are $(\lmd_n,\zp)$. 

One can express $F_{\alp,\alp}(\lmd,z)$, $F_{\alp,\alp-1}(\lmd,z)$ 
and $F_{\alp-1,\alp}(\lmd,z)$ solely in terms of 
the Bessel function of the first kind; 
\bea
 F_{\alp,\alp}(\lmd,z) \eql -\frac{1}{\sin\pi\alp}
 \brc{J_{-\alp}(\lmd)J_\alp(\lmd z)-J_\alp(\lmd)J_{-\alp}(\lmd z)}, 
 \nonumber\\
 F_{\alp,\alp-1}(\lmd,z) \eql \frac{1}{\sin\pi\alp}
 \brc{J_{1-\alp}(\lmd)J_\alp(\lmd z)+J_{\alp-1}(\lmd)J_{-\alp}(\lmd z)}, 
 \nonumber\\
 F_{\alp-1,\alp}(\lmd,z) \eql -\frac{1}{\sin\pi\alp}
 \brc{J_{-\alp}(\lmd)J_{\alp-1}(\lmd z)+J_\alp(\lmd)J_{1-\alp}(\lmd z)}. 
 \label{Frelation}
\eea
The first equation demonstrates that 
$F_{\alp,\alp}(\lmd,z)=F_{-\alp,-\alp}(\lmd,z)$. 
From (\ref{Frelation}), we can see that  
under the exchange~$\alp\exch 1-\alp$, 
\bea
 F_{\alp,\alp}(\lmd,z) & \exch & F_{1-\alp,1-\alp}(\lmd,z) 
   = F_{\alp-1,\alp-1} (\lmd,z) ~, 
 \nonumber\\
 F_{\alp,\alp-1}(\lmd,z) & \exch & -F_{\alp-1,\alp}(\lmd,z) ~. 
 \label{flip_F}
\eea
It follows  from Eqs.(\ref{def_F_Cs}) and (\ref{def_Cs2}) that
\be
 C^{\rm d}_{\alp,n}(\thw) \exch p_{\alp,n}(\thw)C^{\rm s}_{\alp,n}(\thw) ~. 
 \label{flip_C}
\ee

\vskip 2.cm

\def\jnl#1#2#3#4{{#1}{\bf #2} (#4) #3}

\def\Zphys{{\em Z.\ Phys.} }
\def\jssc{{\em J.\ Solid State Chem.\ }}
\def\jpsJ{{\em J.\ Phys.\ Soc.\ Japan }}
\def\ptps{{\em Prog.\ Theoret.\ Phys.\ Suppl.\ }}
\def\PTP{{\em Prog.\ Theoret.\ Phys.\  }}

\def\JMP{{\em J. Math.\ Phys.} }
\def\NPB{{\em Nucl.\ Phys.} B}
\def\NP{{\em Nucl.\ Phys.} }
\def\PLB{{\em Phys.\ Lett.} B}
\def\PL{{\em Phys.\ Lett.} }
\def\PRL{\em Phys.\ Rev.\ Lett. }
\def\PRB{{\em Phys.\ Rev.} B}
\def\PRD{{\em Phys.\ Rev.} D}
\def\PRe{{\em Phys.\ Rep.} }
\def\AP{{\em Ann.\ Phys.\ (N.Y.)} }
\def\RMP{{\em Rev.\ Mod.\ Phys.} }
\def\ZPC{{\em Z.\ Phys.} C}
\def\SCI{\em Science}
\def\CMP{\em Comm.\ Math.\ Phys. }
\def\MPLA{{\em Mod.\ Phys.\ Lett.} A}
\def\IJMPA{{\em Int.\ J.\ Mod.\ Phys.} A}
\def\IJMPB{{\em Int.\ J.\ Mod.\ Phys.} B}
\def\EPJC{{\em Eur.\ Phys.\ J.} C}
\def\PR{{\em Phys.\ Rev.} }
\def\JHEP{{\em JHEP} }
\def\cmp{{\em Com.\ Math.\ Phys.}}
\def\JPA{{\em J.\  Phys.} A}
\def\JPG{{\em J.\  Phys.} G}
\def\NJP{{\em New.\ J.\  Phys.} }
\def\CQG{\em Class.\ Quant.\ Grav. }
\def\ATMP{{\em Adv.\ Theoret.\ Math.\ Phys.} }
\def\ibid{{\em ibid.} }

\renewenvironment{thebibliography}[1]
         {\begin{list}{[$\,$\arabic{enumi}$\,$]}  
         {\usecounter{enumi}\setlength{\parsep}{0pt}
          \setlength{\itemsep}{0pt}  \renewcommand{\baselinestretch}{1.2}
          \settowidth
         {\labelwidth}{#1 ~ ~}\sloppy}}{\end{list}}

\def\reftitle#1{}                

\end{document}